\documentclass[10pt,journal,compsoc]{IEEEtran}
\ifCLASSOPTIONcompsoc
  \usepackage[nocompress]{cite}
\else
  \usepackage{cite}
\fi
\ifCLASSINFOpdf
  \usepackage[pdftex]{graphicx}
  \graphicspath{{./Figures/}}
  \DeclareGraphicsExtensions{.pdf,.jpeg,.png}
\else
  \usepackage[dvips]{graphicx}
  \graphicspath{{./Figures/}}
  \DeclareGraphicsExtensions{.eps}
\fi
\usepackage{amsmath}
\DeclareMathOperator*{\argmin}{argmin}
\usepackage{algorithm}
\usepackage{algorithmic}

\usepackage{array}

\ifCLASSOPTIONcompsoc
  \usepackage[caption=false,font=footnotesize,labelfont=sf,textfont=sf]{subfig}
\else
  \usepackage[caption=false,font=footnotesize]{subfig}
\fi
\usepackage{bm}
\usepackage{ragged2e}
\newcommand{\subparagraph}{}
\usepackage{titlesec}
\titleformat{\paragraph}
{\normalfont\normalsize\slshape}{\theparagraph}{10px}{}
\titlespacing*{\paragraph}
{0pt}{0.8ex plus 1ex minus .2ex}{0.8ex plus .2ex}
\renewcommand\theparagraph{\Alph{paragraph}.}
\usepackage{makecell}
\newcommand\mgape[1]{\gape{$\vcenter{\hbox{#1}}$}}
\begin{document}
%
\title{ART-UP: A Novel Method for Generating Scanning-robust Aesthetic QR codes}
%
%
%
%

\author{Mingliang~Xu, ~\IEEEmembership{Member,~IEEE,}
        Qingfeng~Li, ~\IEEEmembership{Student Member,~IEEE,}
        Jianwei~Niu, ~\IEEEmembership{Member,~IEEE,}
        Xiting~Liu,
        Weiwei~Xu, ~\IEEEmembership{Member,~IEEE,}
        Pei~Lv, ~\IEEEmembership{Member,~IEEE,}
        and~Bing~Zhou,~\IEEEmembership{Member~IEEE}
}

\markboth{Journal of \LaTeX\ Class Files,~Vol.~14, No.~8, August~2015}%
{Shell \MakeLowercase{\textit{et al.}}: Bare Demo of IEEEtran.cls for Computer Society Journals}
%



\IEEEtitleabstractindextext{%
\begin{justify}
\begin{abstract}QR codes are usually scanned in different environments, so they must be robust to variations in illumination, scale, coverage, and camera angles. Aesthetic QR codes improve the visual quality, but subtle changes in their appearance may cause scanning failure. In this paper, a new method to generate scanning-robust aesthetic QR codes is proposed, which is based on a module-based scanning probability estimation model that can effectively balance the tradeoff between visual quality and scanning robustness. Our method locally adjusts the luminance of each module by estimating the probability of successful sampling. The approach adopts the hierarchical, coarse-to-fine strategy to enhance the visual quality of aesthetic QR codes, which sequentially generate the following three codes: a binary aesthetic QR code, a grayscale aesthetic QR code, and the final color aesthetic QR code. Our approach also can be used to create QR codes with different visual styles by adjusting some initialization parameters. User surveys and decoding experiments were adopted for evaluating our method compared with state-of-the-art algorithms, which indicates that the proposed approach has excellent performance in terms of both visual quality and scanning robustness.
\end{abstract}
\end{justify}
\begin{IEEEkeywords}
Aesthetic QR codes, error analysis, visualization optimization, scanning robustness, scanning probability calculation.
\end{IEEEkeywords}}

\maketitle

\IEEEdisplaynontitleabstractindextext

%
\IEEEpeerreviewmaketitle

\IEEEraisesectionheading{\section{Introduction}\label{sec:introduction}}


%
%
%
%

\IEEEPARstart{A}{ quick} response (QR) code is a matrix symbology consisting of an array of nominally square modules. The popularization of smartphones has brought about wide use of QR codes in connection with offline and online life. These codes offer the advantages of large information capacity, low cost, and easy manufacture, etc\cite{ISOIEC200618004}. At the same time, their appearance, consisting of black and white square modules, is difficult to meet the individualized demands of customers because of lack of aesthetic elements. \par
The process of embellishing QR codes always aims to improve the visual quality of their appearance, which makes them more interesting and more appealing\cite{cox2012qart,chu2013halftone,lin2013appearance,zhang2015aesthetic,ramya2015optimized,linefficient}. They can incorporate high-level semantic features such as faces, letters or logos into products of plain design and contribute to brand promotion. However, changing the appearance of QR codes manually is costly and difficult to achieve, as designers have to repeatedly confirm that the result can be recognized by general QR scanners. Consequently, changing in an automatic way is desirable, for the purpose of efficiency and low additional design cost.\par
The main challenge in embellishing QR codes is to ensure that the original information is not influenced and can be scanned by general scanners correctly even when the code's appearance has been changed. Scanning robustness can be defined as the property describing whether the embellished QR code is easy to be scanned correctly by universal scanners. When QR codes are scanned in real-life applications, good scanning robustness is especially important, as it reduces the impact of various environmental factors, such as illumination, noises, coverage, and camera angles, etc, during the scanning process.\par
\begin{figure}[!t]
\centering
\includegraphics[width=3.2in]{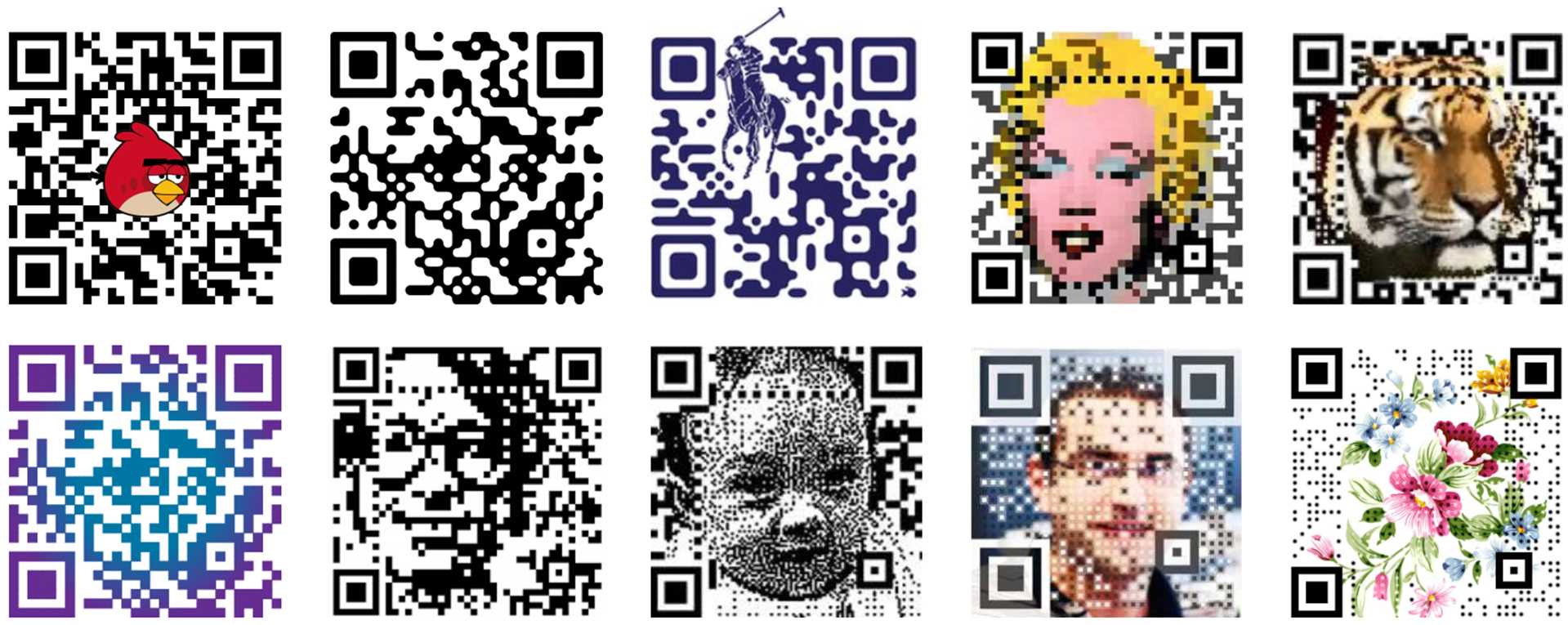}
\caption{Various types of aesthetic QR codes}
\label{fig_variousQRCode}
\end{figure}

Traditional QR code images express information via modules using two highly contrasting colors. While the Reed-Solomon algorithm \cite{macwilliams1977theory} is used to enhance fault tolerance and maintain the scanning robustness\cite{ISOIEC200618004}. On the other hand, embellished aesthetic QR codes enhance visual quality and contain visual information by introducing other colors, changing module shapes, embedding icons, or adjusting codewords. For example, some representative results of aesthetic QR codes are shown in \figurename\ \ref{fig_variousQRCode}. \par
In this work, we propose ART-UP codes - sc\textbf{A}nning-\textbf{R}obust aes\textbf{T}hetic QR (q\textbf{U}ick res\textbf{P}onse) codes. We establish a new module-based scanning probability estimation model to measure the scanning robustness of aesthetic QR codes, resulting in QR codes with the same error tolerance but improved embellishment. More specifically, we first analyze the steps and fundamental reasons that cause errors during QR code scanning in accordance with QR decoding principles. Scanning errors are classified as either thresholding errors or sampling errors. Then, error models are accordingly established, with corresponding estimations of the probability of correctly sampling each module.\par
We then use an iterative luminance adjustment solution to locally adjust the luminance of each module by combining these probability estimation models, so as to increase the probability of correctly sampling and to improve scanning robustness. In addition, in order to obtain better visual quality and make the modification more effective, we combine the image saliency with the corresponding probability constraint, which generates QR codes that are scanning-robust and well embellished.\par
The major contributions of this work are as follows:
\begin{itemize}
\item The scanning and decoding process of QR codes is analyzed, threshold and sampling errors that affect the robustness of embellished QR codes are quantified, and a module-based scanning probability estimation model is built in this paper, which is used to estimate the scanning robustness of the QR code.
\item An optimization strategy based on local luminance adjustment algorithm is proposed, which can balance the tradeoff between visual quality and scanning robustness. And a new method of generating aesthetic QR codes is established, which can achieve state-of-the-art visual quality and preserve error correction abilities.
\item
An interesting threshold estimation algorithm is proposed, which is flexible enough to generate aesthetic QR code images with different styles by adjusting initialization parameters.
\item An easy to implement and effective luminance adjustment algorithm based on linear interpolation is proposed, which is fast, accurate. This algorithm can be applied to convert a grayscale aesthetic QR code into a color one.
\end{itemize}
\section{Related Work}
QR codes were first invented for tracking vehicles and parts during the manufacturing process, recently with the rapid development of mobile Internet, these have been widely applied to many different fields. This has resulted in an increased research interest in the technology necessary for embellishing QR codes\cite{chu2013halftone,cox2012qart,lin2013appearance,garateguy2014qr,fang2014optimization,zhang2015aesthetic,li2015aesthetic,jiang2014qr,baharav2013visually,lin2013artistic,gao2015invisible,yang2016artcode}.\par
Currently, the existing foundation algorithms of generating embellished QR codes can be classified into four groups: embedding icons\cite{samretwit2011measurement}, replacing colors\cite{unitag}, changing the module shape\cite{chu2013halftone,lin2013artistic}, and adjusting codewords\cite{cox2012qart}. Among the four ways, the one that embedding icons is the easiest to implement, but it relies on the inherent error correction capacity of the QR code, so its controllable area is relatively small and monotone, and reduces the error correction ability of the algorithm\cite{samretwit2011measurement,li2015aesthetic}. Replacing colors and changing the module shape for attaching semantic information often need manual intervention and difficult to handle.\cite{lin2013artistic}. Generating QR codes by adjusting codewords usually achieves scanning robustness, but the resulting visual quality is poor and the result is usually inconsistent\cite{cox2012qart}.\par
In recent studies, researchers usually try to mix QR code modules with arbitrary input image in order to obtain a general method. For example, Cox\cite{cox2012qart} looked into the principle of QR code encoding and, by combining the characteristics of Reed-Solomon code with Gauss-Jordan elimination algorithm, proposed a complex method which adjusts the codewords of QR code in the encoding procedure, making the resulting image similar to the binary image. However, this method is fit for encoding URL data only. Through constant optimization, this algorithm has become the most efficient algorithm for adjusting codewords.\par
Apart from this, in 2013 Chu \emph{et~al.}\cite{chu2013halftone} proposed Halftone QR Codes, combining the halftone algorithm with embellished QR codes. By separating each module into 3-by-3 sub-modules and binding the module color to the central sub-module color, a replacement algorithm of was proposed, which offered reliability and regularization in generating halftone QR codes. However, the algorithm adopts a nonlinear optimization procedure which suffers from inefficiency and low-quality visual result issues.\par
Subsequently, Garateguy \emph{et~al.}\cite{garateguy2014qr} proposed a new algorithm using halftone, which was based on the selection of a set of modified pixels using a halftoning mask and a probabilistic model predicting the distortion generated by the embedded image. Although it can enhance the visual quality by sacrificing some scanning robustness, the generated images still contain much image noise.\par
To further improve the visual quality of aesthetic QR codes, Yu-Hsun \emph{et~al.}\cite{lin2013appearance} and Shih-Syun \emph{et~al.}\cite{linefficient} proposed two new algorithms. Yu-Hsun \emph{et~al.} took the saliency the embedded image into consideration so that pixels are modified more effectively, which offers a remarkable improvement in the visual quality of high version QR codes, but the effect of improving low version codes was limited. On the other hand, Shih-Syun \emph{et~al.} proposed an easy and efficient method, where the QR code module sequence was first adjusted to maintain its global similarity with the input image based on the Gauss-Jordan elimination procedure, and then a rendering mechanism was designed to blend the input image into the QR code according to a weight map for a better visual quality. Although the visual effect was quite good, it severely degraded the scanning robustness. \par
Apart from constantly improving the visual effect of QR codes, maintaining scanning robustness is also a challenge that researchers need to take into consideration. Alva \emph{et~al.} \cite{visualead} proposed a commercial algorithm (Visualead) for generating scanning-robust embellished QR codes, but its implementation is closed and its appearance always gives people a feeling of clutter. On the other hand, Zhang \emph{et~al.} \cite{zhang2015aesthetic} came up with an aesthetic QR code generation algorithm based on two-stage image blending, as well as module-based and pixel-based blending. This resulted in a relatively robust embellished QR code generating algorithm, however, this algorithm resorts to empirical rather than theoretical analysis, which results in some images being hard to scan correctly for unexplained reasons, making the algorithm unsuitable for general use.\par
In this paper, we analyze theoretically and guarantee the scanning robustness of QR codes by establishing an error analysis model to estimate the probability of a single module being scanned properly. Meanwhile, by designing an adaptive parametric method, visually appealing and highly robust QR codes are obtained.\par

\section{QR CODE SCANNING}
In order to generate an embellished QR code which is scanning robustness, we must analyze the details of the whole scanning and decoding process, and determine the error-generating factors that may affect scanning. This will be discussed in this section.\par

\subsection{Decoding Algorithm} \label{sec:DecodingAlgorithm}

A QR code is a matrix that consists of white and black modules ordered by a specified encoding rule, where each module's white or black color represents a bit, i.e. a 0 or 1, and 8 bits make up a codeword. Further, the codeword sequence is always subdivided into one or more blocks, and each block is divided into two parts: data codewords and error correction codewords. More information about the generation of QR codes can be found in \cite{ISOIEC200618004}.\par

Typically, the scanning process involves the use of terminal devices to obtain the QR code image from screens or prints via a camera and then using image processing technology to locate and sample the black-and-white modules. According to the sampled information, the QR code image is transformed into a matrix, which is subsequently decoded using a decoding algorithm\cite{owen2013zxing, ohbuchi2004barcode, liu2006automatic, liu2008recognition}.\par
\begin{figure}[!t]
\centering
\includegraphics[width=3.0in]{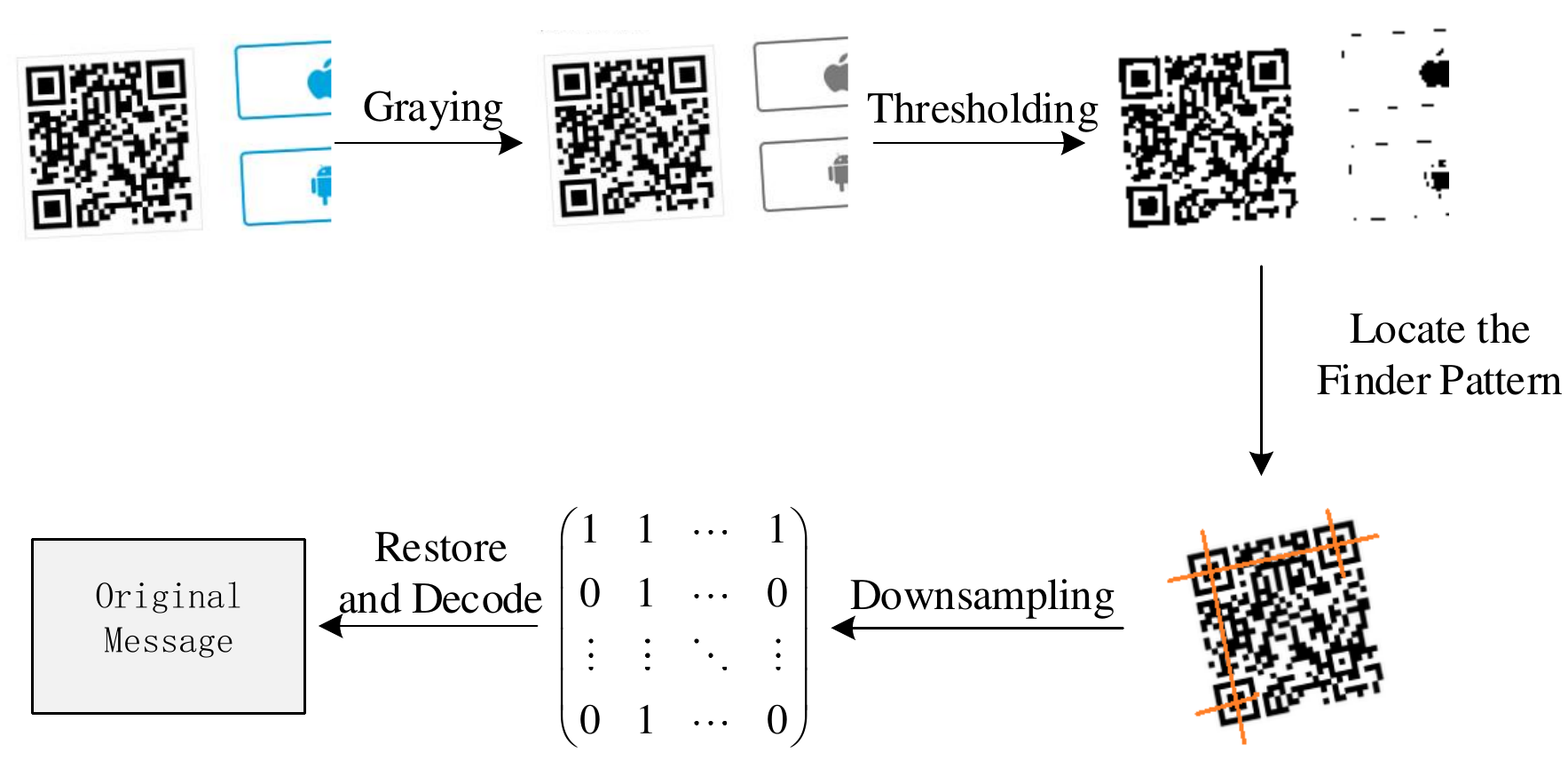}
\caption{QR code scanning process}
\label{fig_scannerQR}
\end{figure}

In accordance with the most well-known QR code generating and scanning approach, based on the open-source library ZXing \cite{owen2013zxing}, we divide the scanning process into three steps: preprocessing, detection and recognition. Preprocessing includes transforming the RGB image into a grayscale one, and then converting the 256-gray intensity level image into a binary one. Detection involves finding and confirming the exact location of QR code, while recognition consists of down-sampling and decoding, as shown in \figurename\ \ref{fig_scannerQR}.\par

\subsubsection{Preprocessing} \label{sec:Preprocessing}
Images captured from the camera and transformed into a binary one during the preprocessing step, which mainly contains two steps. Firstly, converting the RGB image into a single-channel one, using the equation below:
\begin{equation}
\label{eqn_rgb2gray}
{Y_x} = \alpha {C^r_x} + \beta {C^g_x} + \gamma {C^b_x}
\end{equation}
where $\alpha=0.299$, $\beta=0.587$, $\gamma=0.114$, $Y_x$ is the output value of the single-channel image at position of $x$, and $C^r_x$, $C^g_x$, $C^b_x$ respectively correspond the input values of the red, green, and blue channels of the color image at $x$.\par
Then, converting an image into a binary one is also based on the commonly used approach of thresholding:
\begin{equation}
\label{eqn_thresholdComp}
{H_x} = \zeta ({Y_x},{t_x})
\end{equation}
where $H_x$ and $t_x$ are respectively the binary result and threshold at position $x$, $\zeta ({Y_x},{t_x})$ is a function which assigns the value 1 to ${H_x}$ when ${Y_x} \ge {t_x}$, and 0 otherwise.\par
Choosing a proper threshold is key to the binarization effect that influences the performance of QR code detection, and ZXing uses a hybrid local block averaging method for it. That is, the obtained image is first divided into image blocks, the size of each is 8-by-8, without overlapping, and the average of each block is calculated. Then, a 5-by-5 set of blocks is formed around each block and the averages of these blocks are calculated as the final threshold.\par
To facilitate the discussion, in the following, the $i$-th block is denoted as $S_i$, while the set of its neighboring blocks is denoted as ${\mathop{\rm \Gamma}\nolimits} ({S_i})$. When pixel $x$ is located inside block $S_i$, this is written as $x \in S_i$. In the same block, all the pixels will be compared against the same threshold $T_i$. So, for any pixel $x \in S_i$, we have
\begin{equation}
\label{eqn_thresholdCalc}
{t_x} = {T_i} = \frac{1}{{\left| {{\mathop{\rm \Gamma}\nolimits} ({S_i})} \right|}}\sum\limits_{{S_j} \in {\mathop{\rm \Gamma}\nolimits} ({S_i})} {(\frac{1}{{\left| {S_j} \right|}}\sum\limits_{k \in  {S_j}} {{Y_k}} )}
\end{equation}
\par
For general QR codes, this threshold method helps reduce the scanning error, especially for different illumination environments. However, for aesthetic QR codes, the image color will seriously affect the threshold calculation, and it is easy to obtain an unexpected binary result because a too high or too low value of the threshold is calculated. As a result, the reduction of this effect is the key factor for maintaining scanning robustness.\par
\subsubsection{Detection}
In the detection phase, the main task is to locate the position of the QR code from the binary image generated after preprocessing. Pattern matching is used as the primary method for locating the finder pattern. \par
The finder pattern is located on three corners of the QR code image and each component comprises three concentric squares. The three squares are formed by black 7-by-7 modules, white 5-by-5 modules, and black 3-by-3 modules, respectively. During detection, a black-white-black-white-black pattern with corresponding ratios of 1:1:3:1:1, corresponding to the cross-section of the finder pattern, will be matched in order to quickly identify the existence of a QR code. Finally, according to the relative positions of the 3 finder patterns, the exact position and direction of the image are confirmed.\par
\subsubsection{Recognition} \label{sec:recognition}
The process of recognizing QR codes mainly includes sampling and decoding. Sampling refers to first estimating the size of each module, and then obtaining the number of the modules. After the center pixel of each module is sampled to obtain the information of the whole module, finally, a matrix is formed once all the modules have been sampled.\par
On the other hand, the decoding process involves examining the matrix that is obtained during sampling and parsing the contained information. This involves inversely resolving according to the rules of QR encoding, including data masks, codeword rearrangements, error correction, and decoding. The Reed-Solomon algorithm is used in the process of encoding, and is thus necessary for decoding error correction. When the errors in a block are overwhelming, the original data will not be retrievable, resulting in the failure of resolving the QR code.\par
\begin{figure}[!t]
\centering
\includegraphics[width=3.0in]{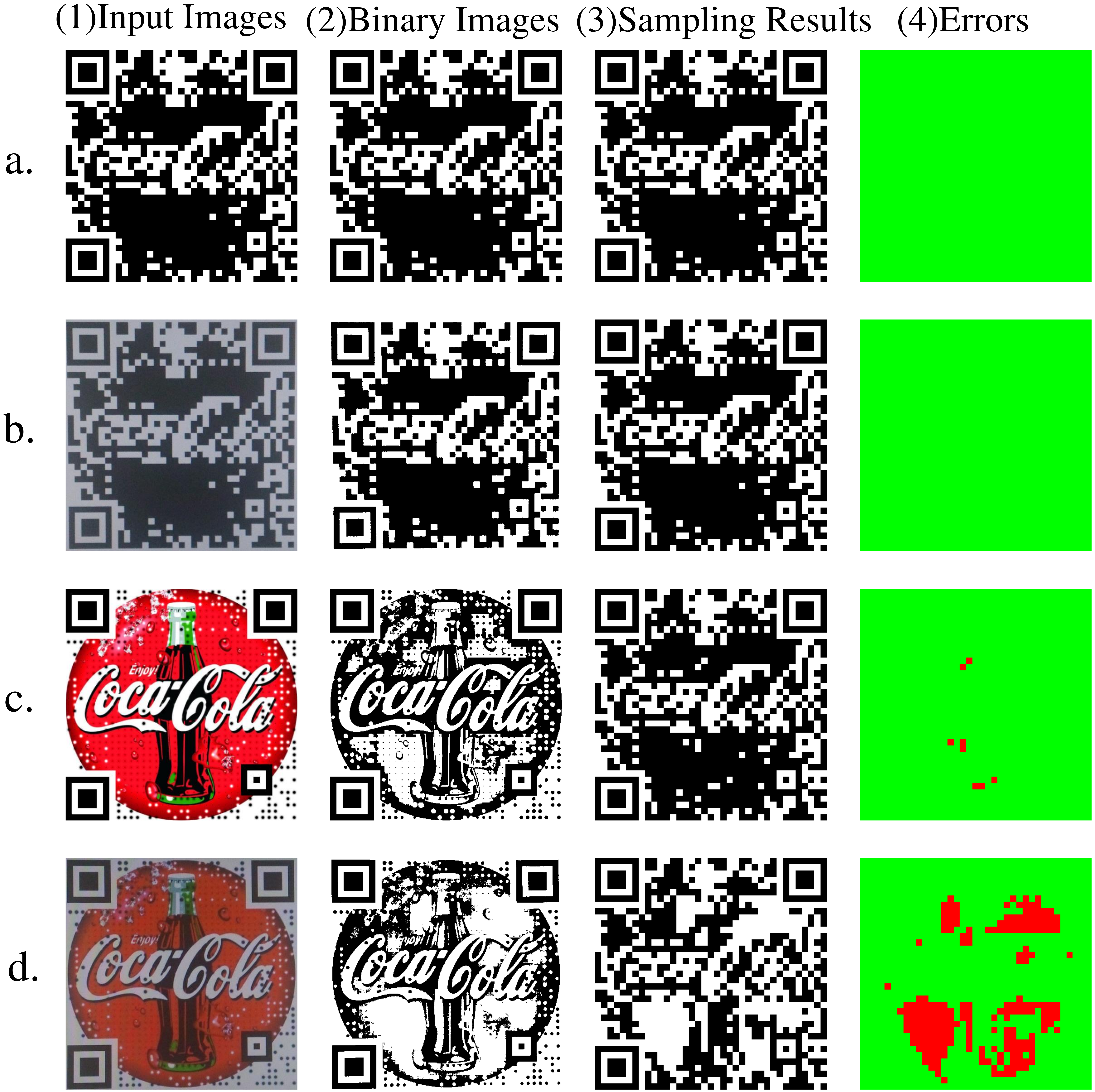}
\caption{Error analysis of the intermediate results of scanning different QR codes using the ZXing library. The last column shows the error images, where errors are depicted in red and correct measurements in green. a. QR code image generated using QART; b. QART QR code image captured using a mobile device under ordinary illumination; c. QR code image generated using CA; d. CA QR code image captured using a mobile device under ordinary illumination.}
\label{fig_qrScannerError}
\end{figure}
\begin{figure*}[!t]
\centering
\includegraphics[width=5.0in]{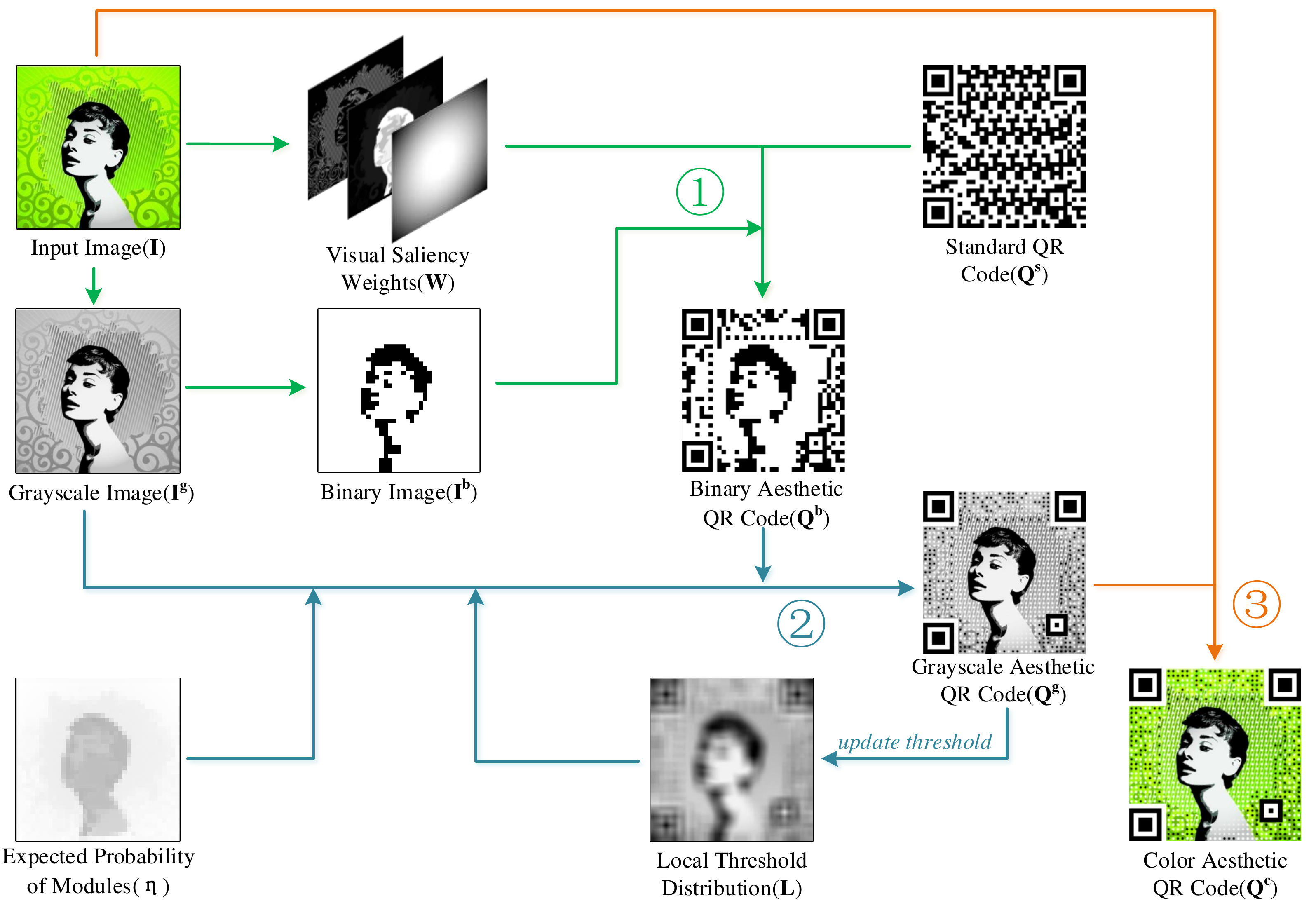}
\caption{Overview of scanning-robust aesthetic QR code generation process. Lines of different colors represent different stages. \textcircled{1} the binary aesthetic stage; \textcircled{2} the grayscale aesthetic stage; \textcircled{3} the color aesthetic stage}
\label{fig_overviewFigure}
\end{figure*}
\subsection{Scanning Error Analysis} \label{sec:ScanningError}
When the QR code is located correctly and the data matrix obtained finally is decoded properly, we regard this code to be an accepted one. However, during actual scanning processes, all kinds of noise and errors may occur. To further analyze and show the various errors produced during the scanning process, we conducted two typical aesthetic QR code implementations for experimentation, QART\cite{cox2012qart} and CA\cite{zhang2015aesthetic}. As shown in \figurename\ \ref{fig_qrScannerError}, we used the ZXing library to compare and maintain the intermediate experimental results for further identification of errors.\par
In \figurename\ \ref{fig_qrScannerError}, column 1 shows the input images. Apart from displaying the scanning process of the original QR code images (Rows (a) and (c)), we used mobile devices to obtain a group of images (Rows (b) and (d)) to compare results with those obtained in conditions of ordinary illumination, i.e. simulating the daily scanning. Column 2 shows the binarization results obtained after processing using the ZXing library. Column 3 displays the downsampling results of QR codes after detection, that is, the matrix of sampling results. In the end, we compare the real sampling result matrix with the expected one, generating Column 4, with red for errors and the green for correctly identified values, which directly shows where errors were generated.\par
As evident from the results, images a-(4) and b-(4) are totally green, which indirectly shows the good scanning robustness of QR codes generated by QART. Regardless of whether they were scanned directly or using mobile devices, the results turned out to be in line with expected ones. However, images c-(4) and d-(4) contain errors, and the errors of d-(4) are much more severe than that of c-(4).\par
According to scanning principles, if there are errors in the matrix of sampling results, they need to be corrected by performing data correction such as using the Reed-Solomon code. When the percentage of the generated errors exceeds a certain threshold, error correction may fail, which results in a scanning failure.\par
This paper proposes a method to estimate the probability of produced errors within a module during the generation process and taking actions to prevent the errors occurring. It obtains the final correct sampling result matrix and improves the scanning robustness, where the details will be described in Section \ref{sec:probSim}.
\section{ART-UP Design and Implementation}
To present the generation of ART-UP codes more clearly, we divide the generation process into three stages: the binary aesthetic stage, the grayscale aesthetic stage, and the color aesthetic stage. In the binary aesthetic stage, codeword adjustment is achieved using visual saliency and Gauss-Jordan elimination, and the order of modules is adjusted to match the binarization result of the input image, which avoids visual conflicts globally. In the grayscale aesthetic stage, thresholding error and sampling error models are established by analyzing the scanning process of general scanners to simulate the process of thresholding and downsampling. In this manner, the probability of correctly sampling each module can be estimated and used as feedback for improving the generation process and ensuring the scanning robustness. In the color aesthetic stage, the pixels of each channel are calculated accurately to create the colored aesthetic QR codes by establishing a linear-based solution, which adjusts the luminance of original images to match the grayscale QR codes.\par
\figurename\ \ref{fig_overviewFigure} shows the whole process of generating ART-UP codes, where $\bf{Q^b}$ is the binary aesthetic QR code, $\bf{Q^g}$ is the grayscale aesthetic one, and $\bf{Q^c}$ is the colored aesthetic one. In the following three Sections, the methods adopted for each stage will be discussed in details.\par
\subsection{Binary Aesthetic QR Codes}
The generation of binary aesthetic QR code is the main pre-step for the generation of grayscale code. The result will directly determine the structural visual effect\cite{wang2004image} of the final colored code. In this step, we first generate a grayscale image $\bf{I^g}$ in accordance with the original one $\bf{I}$ and then generate a binary image $\bf{I^b}$ using a module-based thresholding algorithm. After that, the codewords of the original QR code $\bf{Q^s}$ will be adjusted according to the priority weight $\bf{W}$ in order to generate the target binary aesthetic QR code $\bf{Q^b}$.\par
\subsubsection{Module-Based Binarization}
To make the generated binary aesthetic QR code's similar to the original one, we need to generate a binary image by processing the original QR code. The side length ($n$) of the grayscale image is usually much larger than the length of individual modules ($l = 4 \times {V}+17$, where $V$ is the version of the QR code), so if we only scale the image $\bf{I^g}$ to obtain its binary image, the visual effect is not ideal. Here, the approach used in \cite{zhang2015aesthetic} is adopted to avoid such phenomena. First, the image $\bf{I^g}$ is broken down into several square modules. Let the side length of each such module be $a = \frac{n}{l}$, that is, equal in size to the QR code module. Finally, we mark the $k$-th module as $M_k$, so the binarization of the module is given by
\begin{equation}
{I^b_k} = \zeta (\sum\limits_{x \in {M_k}} {{I^g_x}{G(x)}} ,\frac{{255}}{2})
\end{equation}
where $I^b_k$ stands for the $k$-th element of image $\bf{I^b}$, $G(x)$ is the weight of the pixel $x$, and $\sum\limits_{x \in {M_k}} {G(x)} { = }1$. In this paper, $G(x)$ is a Gaussian Distribution:
\begin{equation}
\label{eqn_gaussianBlockWeight}
{G(x)} = {G_{{M_k}}}(i,j) = \frac{1}{{2\pi {{{\sigma}_1} ^2}}}{e^{\frac{{ - {(i-a/2)^2} + {(j-a/2)^2}}}{{2{{{\sigma}_1} ^2}}}}}
\end{equation}
$x$ belongs to $M_k$, and ${G_{{M_k}}}(i,j)$ refers to the weight of the $k$-th module at the position of $(i,j)$. It is necessary to note that in Equation \ref{eqn_gaussianBlockWeight}, the position of the $k$-th module $(i,j)$ corresponds to the global position of $x$. In addition, in our experiment, we set ${{\sigma}_1}$ to be $\frac{a-1}{5}$.\par
\subsubsection{Codewords Adjustment}
From the analysis of QR code scanning principles, it is evident that the scanned QR code will be accepted by the scanners as long as the order of the modules passes the error correction stage. QR codes employ Reed-Solomon encoding, which has the following properties: 1) It is a systematic code, where the input data is embedded in the final encoded output. That is, the first half is the original input code, with the latter half being the error correction code; 2) The result of an exclusive-or between two different Reed-Solomon encoded blocks, is also a valid Reed-Solomon encoded block.\par
During encoding, each block is an independent Reed-Solomon code, and a block can be separated into 3 areas. One is for input data bits, one for padding bits, and the rest for correction bits, with the corresponding lengths being $m$, $p$, and $c$, respectively. According to the QR encoding rules, when the version and error correction level of a QR code are confirmed, $m+p$ and $c$ are fixed constants, while $m$ and $p$ may change according to the size of the input data.\par
\begin{figure}[!t]
\centering
\subfloat[]{
\includegraphics[width=2.9in]{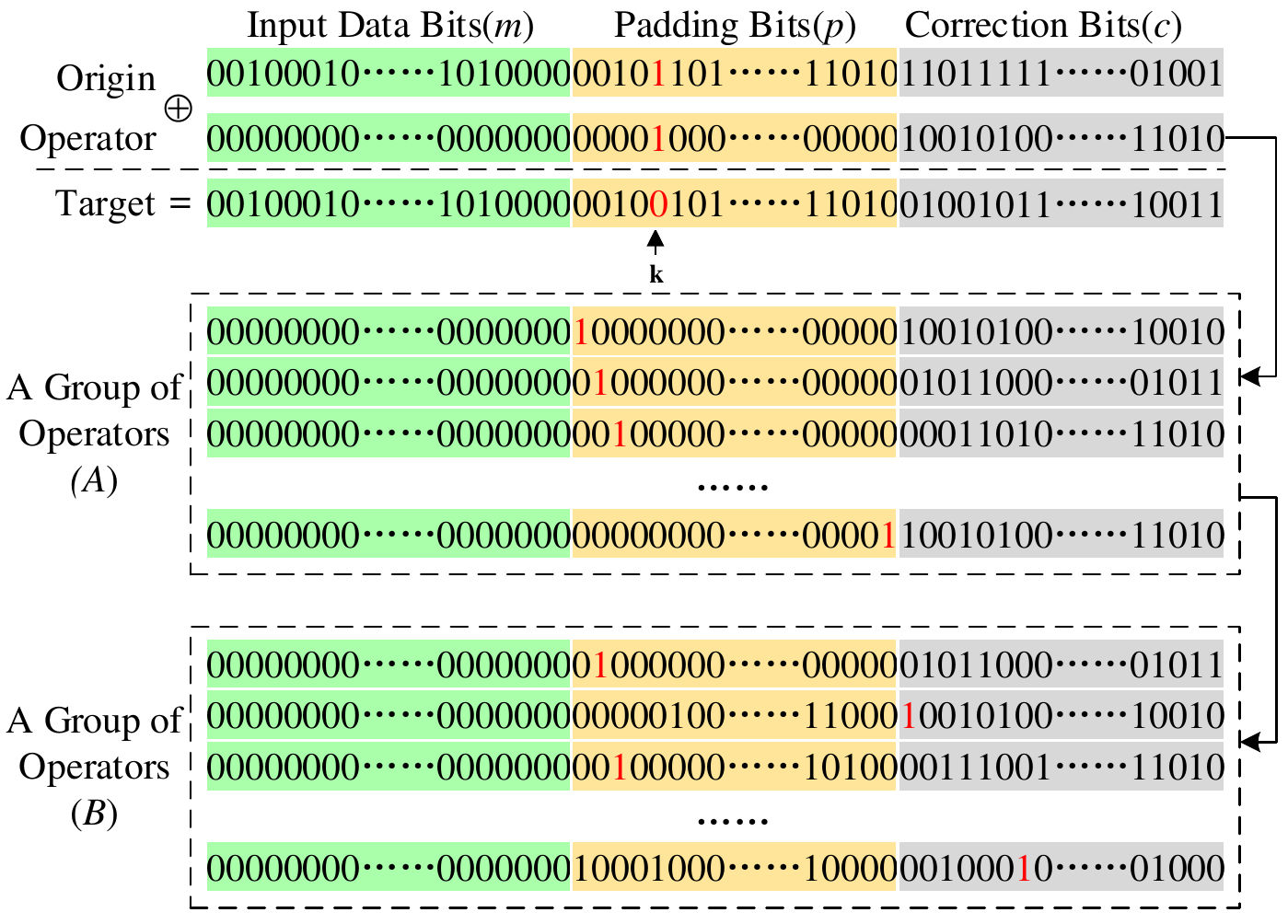}
\label{fig_coxOperate}
}
\hfil
\subfloat[]{
  \includegraphics[width=0.85in]{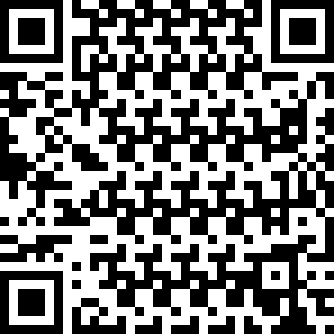}
  \label{fig_coxOrigin}
}
\hfil
\subfloat[]{
  \includegraphics[width=0.85in]{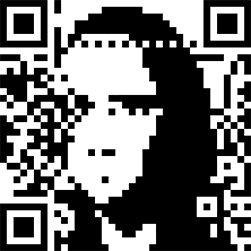}
  \label{fig_coxSetA}
}
\hfil
\subfloat[]{
  \includegraphics[width=0.85in]{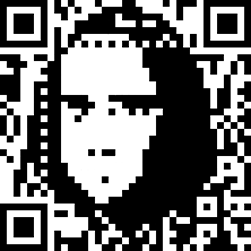}
  \label{fig_coxSetB}
}
\caption{Codeword adjustment using the Gauss-Jordan elimination method.
a. Generation of the operator set;
b. Original QR code;
c. Binary result of codeword adjustment by operator set A;
d. Binary result of codeword adjustment by operator set B;
}
\label{fig_coxOperatorExample3}
\end{figure}

To describe the process of codeword adjustment clearly, consider \figurename\ \ref{fig_coxOperate} as an example. The first row contains the original information. In order to generate the aesthetic binary QR code $\bf{Q^b}$ in \figurename\ \ref{fig_overviewFigure} similar to the binary image $\bf{I^b}$, let us assume that the $k$-th ($k>m$ and $k \le m+p$) digit needs to be adjusted from 1 to 0 in the block. Then, a special operator is constructed whose $k$-th digit is 1 while the other digits in the input data and padding bits are 0. The correction bits are generated by the checking rules of Reed-Solomon coding. According to Property 2, a legal Reed-Solomon code block will be obtained through the exclusive-or operation of the special operator and the original code. The generated code will maintain the other digits of the input data area and padding area unchanged, while the $k$-th digit will go through an inversion operation.\par
As described above, a group of operators $A$ can be constructed, as shown in \figurename\ \ref{fig_coxOperate}. There are $p$ operators marked as ${a_1}, {a_2}, \cdots, {a_p}$, respectively, where digits $m+1$, $m+2$, $\cdots$, $m+p$ of the corresponding operators are 1, while the rest of the digits of the input data area and padding area are 0. That is, there is only one row with a 1 for each controllable module in the padding data area, which is marked with red. With this group of operators, we can apply an exclusive-or operation on the current data with $a_{k-m}$ to manipulate the $k$-th digit to the desired value without affecting any of the other controllable modules. The intermediate result adjusted by Set $A$ is shown in \figurename\ \ref{fig_coxSetA}. \par
In order to maintain the input data bits unchanged and the error correction bits updated, the controllable modules must be limited within the padding data area ($k>m$ and $k \le m+p$). However, the Gauss-Jordan elimination method can be adopted to overcome these constraints and allow further flexibility. By combining the operators in set $A$, we can create a new operator set $B$ and apply an exclusive-or between the operators to create new basis operator that trades data bits for error correction bits. In this way, the uncontrollable modules are dispersed, and the noise pixels are distributed over the canvas, as shown in \figurename\ \ref{fig_coxSetB}.\par
It should be noted that the input data bits (including the ending indicator) have not been changed during this process, so according to Property 1, the information contained in the QR code is not affected by the codewords' adjustment, which is necessary to ensure a correct decoding outcome. \par
\subsubsection{Visual Saliency Optimization}
As known above, during codeword adjustment, the controllable modules can be only moved within a certain range due to the QR codes' encoding rules' limitation, no new modules can be added. In each block, the color of the $p$ controllable modules can be freely controlled, while the other areas will be affected by the input data and the checking algorithm. Therefore, a priority algorithm is proposed on how to choose controllable modules.\par
In this paper, a linear combination of saliency, edge detection, and heuristic constraints is used to determine this priority for each module of QR code:
\begin{equation}
{\bf{W}} = {\lambda _1}{\bf{Edge}} + {\lambda _2}{\bf{Sal}} + {\lambda _3}{\bf{Heu}}
\end{equation}
${\bf{Edge}}$ and ${\bf{Sal}}$ are the results of edge detection and visual saliency respectively for image $\bf{I}$. To make the $\bf{W}$ and $\bf{Q^s}$ equal in size, a mean-pooling operation is carried out. ${\bf{Heu}}$ defines some heuristic rules, so that in cases where two modules have similar ${\bf{Edge}}$ and ${\bf{Sal}}$, it allows the one which is closer to the image center to obtain a larger weight.
\begin{equation}
\begin{split}
Heu_{x,y} &= 1 - \frac{{{{(x - \frac{1}{2}l)}^2} + {{(y - \frac{1}{2}l)}^2}}}{{\frac{1}{2}{l^2}}} \\
  &= \frac{{xl + yl - {x^2} - {y^2}}}{{\frac{1}{2} {l^2}}} \\
\end{split}
\label{eqn_heu_var}
\end{equation}
In Equation \ref{eqn_heu_var}, $Heu_{x,y}$ refers to the heuristic variable's result at the module position $(x,y)$. $l$ is the number of the modules on each side of the code, that is, $l = 4V+17$. In this paper, the Canny \cite{canny1986computational} and Region-based Contrast (RC) \cite{cheng2015global}  methods are adopted for the edge detection and saliency area extraction, respectively. After normalizing ${\bf{Edge}}$, ${\bf{Sal}}$ and ${\bf{Heu}}$, the values of $\lambda _1$, $\lambda _2$, and $\lambda _3$ are determined to be 0.67, 0.23 and 0.10 respectively in our experiments. Although ${\bf{Heu}}$ only has a small weight, for most of images provided by users, such as logos and human faces, the choice of controllable modules always plays a key role in the final processing.\par
\subsection{Grayscale Aesthetic QR Codes}
To further enrich the details of QR code, the visual effect needs to be optimized by combining it with an image to generate a grayscale one. As shown in \figurename\ \ref{fig_overviewFigure}, first the initial local threshold distribution ${\bf{L}}$ according to the grayscale image $\bf{I^g}$ and binary aesthetic QR code $\bf{Q^b}$ is pre-estimated. Then, a module-based luminance modification algorithm is used to adjust the pixels of the modules in the corresponding grayscale image $\bf{I^g}$, ensure scanning robustness in accordance with the expected probability distribution $\bm{\eta}$, and combine it with $\bf{Q^b}$ to generate a grayscale QR code $\bf{Q^g}$. After that, the local threshold distribution ${\bf{L}}$ of the real image is recalculated and updated through $\bf{Q^g}$, and luminance modification is used to recreate $\bf{Q^g}$. With continuous iterations that make the local threshold distribution gradually steady, the final $\bf{Q^g}$ is obtained.\par
During generation of grayscale aesthetic QR codes, a module-based scanning probability estimation model is proposed in this paper, which is used to estimate the scanning robustness of the QR code. After that, an iterative image color updating algorithm is proposed to adjust the grayscale aesthetic QR code, making it robust and visually appealing. These will be introduced in the following sections.\par
\subsubsection{Module-based Scanning Probability Estimation} \label{sec:probSim}
According to the scanning process analyzed in Section \ref{sec:DecodingAlgorithm} and the scanning error analysis of Section \ref{sec:ScanningError}, we classify errors resulting from scanning failure into two types: thresholding errors and sampling errors. Thresholding errors mainly refer to those that result from the difference between the real and expected threshold due to the environment, camera devices, and image content local thresholding. Sampling errors occur during sampling because of location and scaling errors, where the real sampling position is far from the module center. In the following, we will analyze these in detail and propose the module-based scanning probability estimation model.\par
\paragraph{Thresholding Error Estimation}
As can be inferred from Section \ref{sec:Preprocessing}, generic QR code scanners use the thresholding method based on hybrid local block averages. In theory, the threshold applied to a certain pixel only depends on image blocks and their neighboring pixels during the thresholding process. However, thresholding problems are complicated in reality, the following factors must be taken into account: 1) In most realistic scenarios, the colors of the obtained pictures may differ from the original ones as they are affected by environmental conditions; 2) In the thresholding method based on the hybrid local block averages, the captured image is divided into non-overlapping image blocks at first, the adapted thresholds are the same within the same block, but actually the position of the pixels cannot be confirmed during separation because of the unknown position and angle of the obtained QR code in the image; 3) In theory, the size of the image blocks is defined accurately, but in reality due to the scanning distance and image scaling, the relative size of the image block and the QR code module is not fixed.\par
Although an accurate determination of actual threshold values of pixels may be difficult to achieve, good estimations may be possible under certain assumptions. Firstly, as for the problem of color deviation, usually the captured color is similar to the real color. Without considering the limitations of color thresholding, we can suppose the probabilities that the captured color is more or less intense than the real one are equal, and with the increase of the changing amplitude, these probabilities are gradually reduced. Secondly, due to the randomness of the position and angle of the code within an image, resulting in random pixel locations for the blocks, we may as well assume that the pixel is always located in the center of the block. In this way, in accordance with Equation \ref{eqn_thresholdCalc}, the thresholding method based on hybrid local block averages can be replaced by the thresholding method based on the local average, that is, the average of the neighboring pixels can be used to determine the threshold. Luckily, even if the pixel is not located in the block's center, its real threshold can be estimated from the surrounding area assuming that the local image area exhibits color continuity. In addition, the influence of the scanning distance and image scaling have on the real threshold is random, and depends on the scanning performed by the users, resulting in an unknown percentage of the QR code lying within the image, usually clustered within a certain area, which means that the threshold to change within the corresponding ranges.\par
According to the above analysis and observations, it is evident that for the QR code, the same pixel under different environments may be compared to different thresholds, resulting in different binarization results. General methods of generating aesthetic QR codes are usually applied with the assumption that the thresholds are constant. This assumption leads to a reduction in scanning robustness. In this paper, in order to simulate the real thresholding process, we assume that the expected threshold of a certain pixel $x$ in an aesthetic QR code is obtained from the average of its surrounding pixels, denoted as $t_{xo}$. The actual threshold value, which is affected by the environment, lies close to this value, and follows a Gaussian distribution ${{p_\tau }(t_x)}$, with $t_x=t_{xo}$ being its maximum value.\par
\begin{equation}
\label{eqn_thresholdDistribution}
{p_\tau }({t_x}) = \frac{1}{{\sqrt {2\pi } {{\sigma}_2} }}{e^{\frac{{ - {{({t_x} - {t_{xo}})}^2}}}{{2{{{\sigma}_2} ^2}}}}}
\end{equation}
Meanwhile, according to the above, we have $t_{xo}$
\begin{equation}
\label{eqn_predictThreshold}
t_{xo} = \frac{1}{{\left| {{\mathop{\rm R}\nolimits} (x)} \right|}}\sum\nolimits_{i \in {{\mathop{\rm R}\nolimits} (x)}} {{Y_i}}
\end{equation}
, Where ${\mathop{\rm R}\nolimits} (x)$ represents the set of pixels neighboring the center $x$. In this paper, ${\mathop{\rm R}\nolimits} (x)$ is a square area, with the side length of $3a$, where $a$ is the side length of the QR module.\par
When we obtain the distribution of thresholds in position $x$, we compare the original pixel value with the real threshold of this pixel, according to Equation \ref{eqn_thresholdComp}. In the meantime, due to the limitations of color thresholding, the real threshold is actually distributed in the range of $[0,255]$. Therefore,
the probability that the thresholding result equals 0 or 1 is, respectively
\begin{eqnarray}
\label{equ_thresholdEstimate1}
P({H_x} = 1) &= P({Y_x} \ge {t_x}) &= \int_0^{{Y_x}} {{p_\tau }({t_x})} \mathrm{d} t_x\\
\label{equ_thresholdEstimate2}
P({H_x} = 0) &= P({Y_x} < {t_x}) &= \int_{{Y_x}}^{255} {{p_\tau }({t_x})} \mathrm{d} t_x
\end{eqnarray}\par
Similarly, because of the limitations of color thresholding, the sum of the two equations above will not be equal to 1. In this paper, it is normalized for adjustment, so when the pixel $x$ is located in the module $M_k$, that is, $x \in M_k$, the probability of correctly thresholding the pixel is
\begin{equation}
\label{eqn_thresholdProbability}
\setlength{\nulldelimiterspace}{0pt}
p^t_x = {{\mathop{\rm p}\nolimits} ^t}({Y_x}) =\left\{\begin{IEEEeqnarraybox}[\relax][c]{l?s}
{\frac{P({H_x}=1)}{{P({H_x}=0)}+{P({H_x}=1)}}},&for $Q^b_k=1$\\
{\frac{P({H_x}=0)}{{P({H_x}=0)}+{P({H_x}=1)}}},&for $Q^b_k=0$%
\end{IEEEeqnarraybox}\right.
\end{equation}\par
\begin{figure}[!t]
\centering
\includegraphics[width=2.4in]{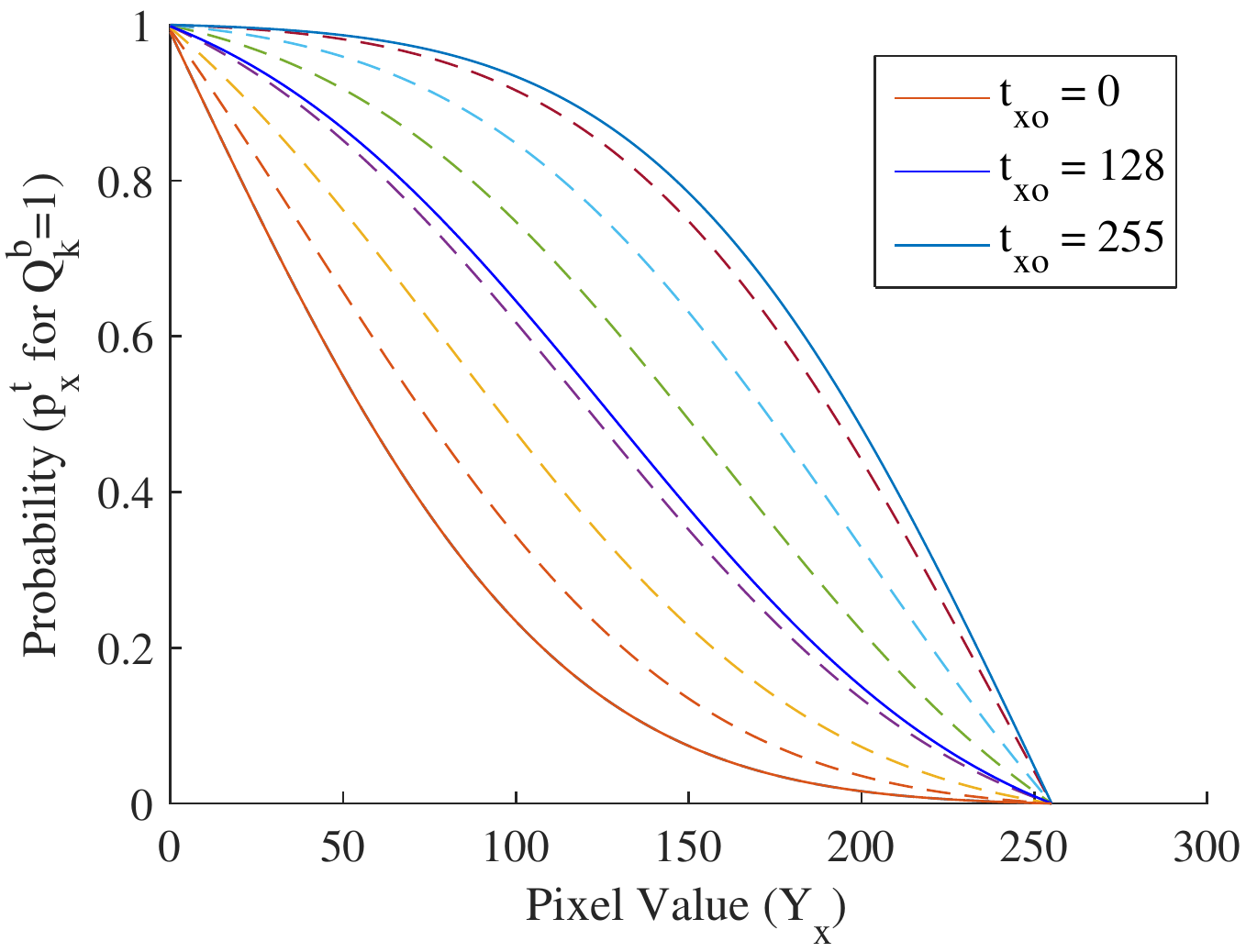}
\caption{Variation curve of $p^t_x$ with $Y_x$ corresponding to different values of $t_{xo}$ when the color of corresponding module is black for a binary aesthetic QR code ($Q^b_k=1$).}
\label{fig_thresholdProbability}
\end{figure}

In \figurename\ \ref{fig_thresholdProbability}, the $Y_x$-$p^t_x$ curves under different values of $t_{xo}$ are shown for ${\sigma}_2 = \frac{255}{3}$, $Q^b_k=1$. From this, it can be derived that when the expected threshold of the pixel is fixed, the probability of correctly thresholding the pixel is monotonically related to the real color luminance, that is, we can adjust the grayscale value of a pixel to raise the probability and thus increase robustness.\par
\paragraph{Sampling Error Estimation}
According to Section \ref{sec:recognition}, in ideal conditions, the downsampling results only relate to the center pixel of the module in the scanning process. However, in real environments, images captured by cameras are likely to be different from the original ones, due to rotations, scaling, and even transfiguring, which result in the sampling error being affected by the surroundings.\par
\begin{figure}[!t]
\centering
\includegraphics[width=2.0in]{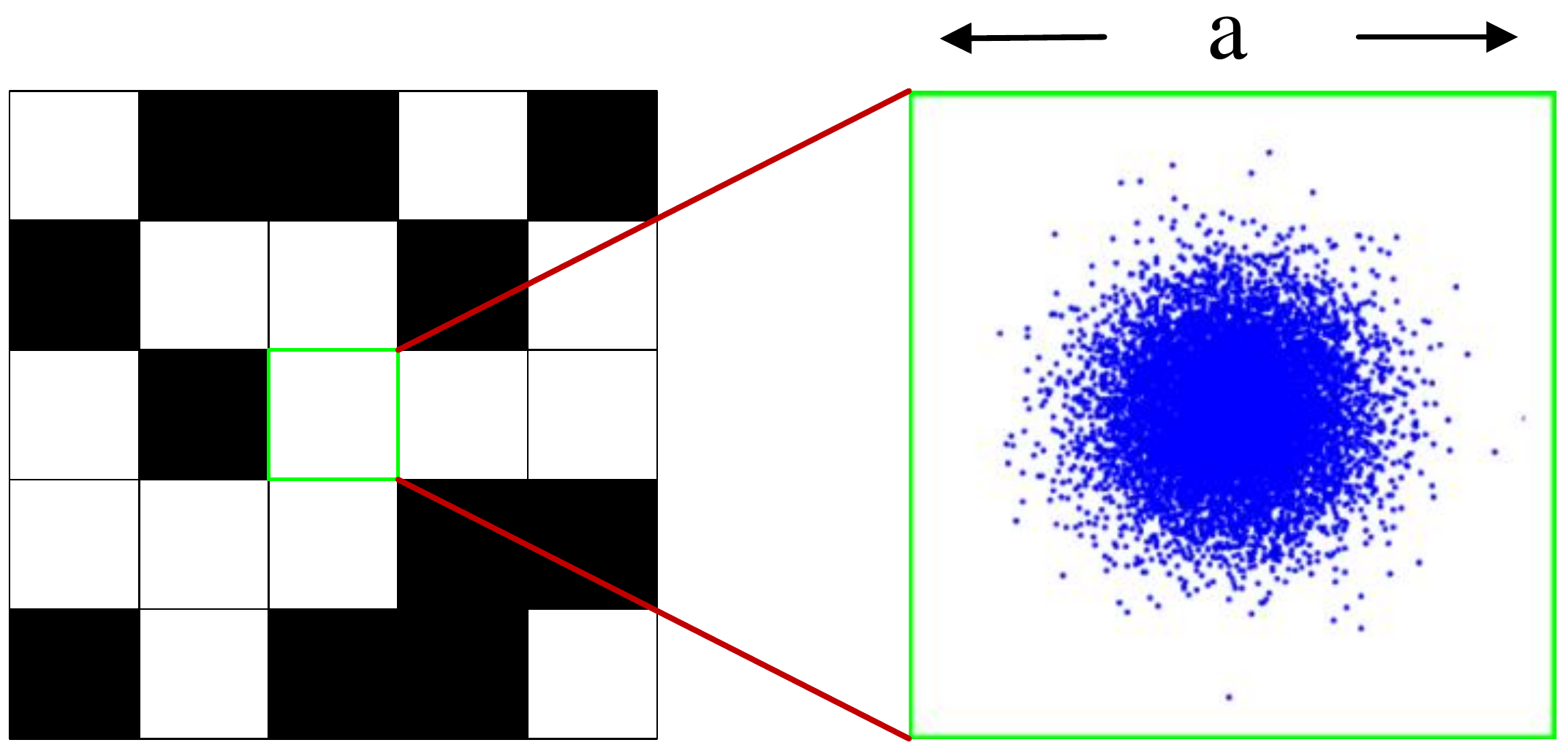}
\caption{The probability distribution of pixels being sampled within a module. We assume that the module is sampled one thousand times, and each sample corresponds to a blue point.}
\label{fig_sampleProbability}
\end{figure}

Furthermore, this sampling error can hardly be simulated realistically, as this would require different application surroundings and large amounts of experiments. In this paper, a reasonable assumption is made that the probability of a pixel being sampled in a module follows a Gaussian distribution. As shown in \figurename\ \ref{fig_sampleProbability}, in a module, the closer the point is to the center, the more likely it will be sampled.
\begin{equation}
\label{eqn_sampleProbability}
{p^s_x} = {{p^s}_{M_k}}(i,j) = \frac{1}{{2\pi {{{\sigma}_3} ^2}}}{e^{-\frac{{ {({i-a/2})^2} + {({j-a/2})^2}}}{{2{{{\sigma}_3}^2}}}}}
\end{equation}
In addition, in the whole module, the sum of the probabilities for sampled pixels is 1, that is, $\sum\limits_{x \in M_k}{p^s_x}=1$.\par
\paragraph{Probability of Correctly Scanning a Module }
As can be inferred from the scanning process analysis, thresholding and downsampling are performed during different steps of the process, and the result of thresholding a pixel is independent of sampling. In this way, a pixel will be selected as the sampling result according to the sampling process, so the probability of sampling this pixel as the correct result is:
\begin{equation}
\label{eqn_pointProbability}
{p^m_x} = {p^s_x}{p^t_x}
\end{equation}
Therefore, the probability of correctly scanning a module is
\begin{equation}
\label{eqn_moduleProbability}
{P_{M_k}} = \sum\limits_{x \in M_k}{p^m_x}
\end{equation}
\subsubsection{Luminance Adjustment}
To balance the scanning robustness and the visual quality of aesthetic QR code, in this section, a luminance modification algorithm within the module is proposed to reconcile the binary aesthetic QR code $\bf{Q^b}$ and the grayscale image $\bf{I^g}$. In addition, the expected threshold of the pixels is established by means of an iterative algorithm. In this manner, the grayscale aesthetic QR code $\bf{Q^g}$ is finally obtained.\par
\begin{algorithm}[!t]
\algsetup{linenosize=\footnotesize}
\footnotesize
  \caption{Luminance Modification Algorithm}
  \label{alg_adjustPointProbability}
  \begin{algorithmic}[1]
    \REQUIRE $Y_x$,$t_{xo}$,$p^s_x$,${{\varpi}^p_x}$ for ${x \in M_k}$, and ${{\eta}_k}$
    \ENSURE ${P_{M_k}} \ge {{\eta}_k}$ in module $k$ of $\bf{Y}$
    \STATE Calculate $p^t_x$ for $x \in M_k$, see equations \ref{eqn_thresholdDistribution},\ref{equ_thresholdEstimate1},\ref{equ_thresholdEstimate2},\ref{eqn_thresholdProbability}
    \STATE $maxpt \gets 1$
    \WHILE {\TRUE}
      \STATE ${P_{M_k}} \gets \sum\limits_{x \in M_k}{{p^s_x}{p^t_x}}$
        \IF {${P_{M_k}} \ge {{\eta}_k}$}
          \STATE \textbf{break}
        \ENDIF
        \STATE $\wp \gets {{\sum\limits_{x \in M_k} {p^s_x} {{\varpi}^p_x}}}$
        \IF {$\wp = 0$}
          \STATE \textbf{break}
        \ENDIF
        \FOR {each $x \in M_k$}
          \STATE $p^t_x \gets p^t_x + \frac{{\left[ {\max({{\eta}_k} ,P_{M_k}) - P_{M_k} } \right]{\varpi}^p_x}}{\wp}$
          \IF {$p^t_x \ge maxpt$}
            \STATE $p^t_x \gets maxpt$
            \STATE ${{\varpi}^p_x} \gets 0$
          \ENDIF
        \ENDFOR
    \ENDWHILE
    \FOR {each $x \in M_k$}
      \STATE find $Y_x^* = \mathop{\argmin}\limits_{Y_x} {\left|{{\mathop{\rm p}\nolimits} ^t}({Y_x}) - p^t_x\right|} $
      \STATE $Y_x \gets Y_x^*$
    \ENDFOR
  \end{algorithmic}
\end{algorithm}
\paragraph{Luminance Modification Algorithm}\label{sec:LuminanceAdjustment}
The whole QR code is made up of many modules. 
By estimating the probability of correctly scanning a single module, we can indirectly work out and adjust the error probability in the matrix of the sampling result. To guarantee the code's scanning robustness, this probability must be modified accordingly so as to be able to recover from errors that may occur through error correction.\par
In this section, we assume that the expected threshold ${t}_{xo}$ of each pixel during the thresholding process is a known variable. So, according to Equation \ref{eqn_thresholdProbability}, the corresponding $p^t_x$ can be calculated. Also, in accordance with Section \ref{sec:probSim}, the probability $P_{M_k}$ of correctly scanning each module in the grayscale aesthetic QR code can be calculated, under the constraint of correctly scanning the modules in order to maintain the scanning robustness of the whole QR code,
\begin{equation}
{P_{M_k}} \ge {{\eta}_k}
\end{equation}
where ${{\eta}_k}$ is the minimum probability constraint for the expected correct scanning of the $k$-th module. Here, an initial fixed value can be given, such as $75\%$. The choice of this value will be discussed later in Section \ref{sec:parameterAnalysis}.\par
As can be inferred from Equation \ref{eqn_sampleProbability}, after confirming the size and version of the QR code image, the probability $p^s_x$ of each pixel in the module is actually a fixed value. So, according to Equation \ref{eqn_pointProbability} and \ref{eqn_moduleProbability}, when ${P_{M_k}} < {{\eta}_k}$, where the probability of correctly scanning the module is lower than the expected one, we have to adjust ${P_{M_k}}$ to meet the constraint by increasing the probability $p^t_x$ of correctly thresholding the pixels in the module. In order to achieve this, the following equation is formed for adjusting the thresholding probability within a module:
\begin{equation}
\label{eqn_adjustPointProbability}
p^t_x \gets p^t_x + \frac{{\left[ {\max({{\eta}_k} ,P_{M_k}) - P_{M_k} } \right]{\varpi}^p_x}}{{\sum\limits_{x \in M_k} {p^s_x} {{\varpi}^p_x}}}
\end{equation}
where ${\varpi}^p_x$ is the weight for adjusting the thresholding probability at position $x$. It is not difficult to see that when the equation is multiplied by $p^s_x$, and all the elements within the module ${M_k}$ are summed, $P_{M_k} \gets P_{M_k}+{\left[ {\max({\eta}_k,P_{M_k}) - P_{M_k}}\right]}$ can be maintained. In this manner, the probability of scanning the module can be updated to become larger than or equal to ${{\eta}_k}$.\par
However, Equation \ref{eqn_adjustPointProbability} ignores the limitation of $p^t_x$, that is, $0 \le p^t_x \le 1$. During the actual updating process, this may result in a phenomenon where the updated probability of correctly scanning the module may be smaller than expected. Therefore, a simplified iterative updating algorithm is proposed (Algorithm \ref{alg_adjustPointProbability}). During the iterations, ${{\varpi}^p_x}$ is adjusted dynamically according to $p^t_x$, thus solving the above problems.\par
Through this algorithm, the correct thresholding probability of each pixel in an image can be assigned to acquire $p^t_x$ and make the probability of correctly scanning each module meet the expected value. As evident from \figurename\ \ref{fig_thresholdProbability}, the grayscale value $Y_x$ that each pixel adopts reflects the probability of correctly thresholding pixel ${p^t_x}$ when $t_{xo}$ is fixed. So, by looking up in the table, the grayscale value corresponding to $p^t_x$ can be easily obtained and the luminance modification of the pixels in an image can be accomplished. \par
\begin{algorithm}[!t]
\algsetup{linenosize=\footnotesize}
\footnotesize
  \caption{Threshold Estimation Algorithm}
  \label{alg_thresholdEstimationAlgorithm}
  \begin{algorithmic}[1]
      \REQUIRE ${{\varpi}^p_x}$ for each $x$ and ${{\eta}_k}$ for each $k$
      \ENSURE Grayscale aesthetic QR code $\bf{Q^g}$
      \STATE Calculate ${p^s_x} \gets {{p^s}_{M_k}}(i,j) = \frac{1}{{2\pi {{{\sigma}_3} ^2}}}{e^{-\frac{{ {({i-a/2})^2} + {({j-a/2})^2}}}{{2{{{\sigma}_3}^2}}}}}$ for each $x$
      \STATE Initialize $\bf{Q^g} \gets 0.5 \times \bf{I^g} + 0.5 \times \bf{Q^b}$
      \STATE Initialize ${{\bf{L}}^{{\bf{old}}}}$,${\bf{L}}$,$n \gets 0$
      \WHILE {\TRUE}
        \STATE $\bf{Y} \gets \bf{Q^g}$
        \FOR {each $x$}
          \STATE $t_{xo} \gets \frac{1}{{\left| {{\mathop{\rm R}\nolimits} (x)} \right|}}\sum\nolimits_{i \in {\mathop{\rm R}\nolimits} (x)} {Y_i}$
          \STATE $L_x \gets t_{xo}$
        \ENDFOR
        \IF {$n\ne 0 $}
          \IF {${{\bf{L}}^{{\bf{old}}}}={\bf{L}}$}
            \STATE \textbf{break}
          \ENDIF
        \ENDIF
        \STATE $n \gets n+1$
        \STATE $\bf{Y} \gets \bf{I^g}$
        \FOR {each $M_k$}
          \STATE prepare $Y_x$,$L_x$ as $t_{xo}$,$p^s_x$,${{\varpi}^p_x}$ for ${x \in M_k}$, and ${{\eta}_k}$
          \STATE update each $Y_x$ using algorithm \ref{alg_adjustPointProbability} for $x \in M_k$
        \ENDFOR
        \STATE $\bf{Q^g} \gets \bf{Y}$
        \STATE $\bf{L}^{\bf{old}} \gets \bf{L}$
      \ENDWHILE
  \end{algorithmic}
\end{algorithm}
\paragraph{Threshold Estimation Algorithm}
In Algorithm \ref{alg_adjustPointProbability}, it was assumed that the adopted expected threshold ${t_{xo}}$ of each pixel is known, and the luminance of the pixels was adjusted based this threshold. However, according to Equation \ref{eqn_predictThreshold}, ${t_{xo}}$ is related to luminance ${Y_x}$. So, when adjusting the luminance of the pixels, the expected threshold will be changed.\par
To find a proper threshold which makes the adjusted luminance of the pixels in line with the assumed expected threshold, an iterative threshold estimation algorithm (Algorithm \ref{alg_thresholdEstimationAlgorithm}) is developed, which estimates the expected threshold and gradually converges to the desired value.\par
\begin{figure}[!t]
\centering
\includegraphics[width=3.0in]{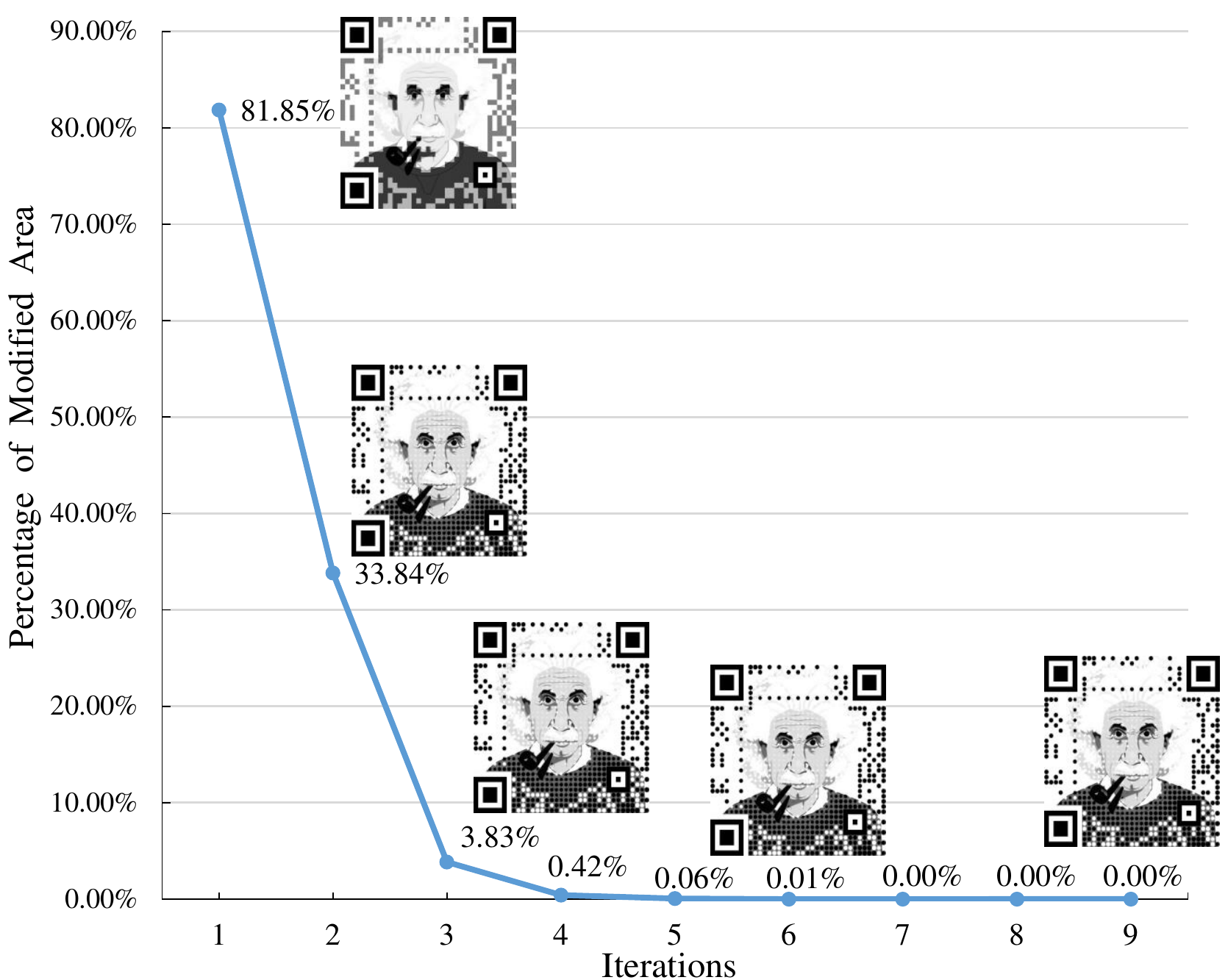}
\caption{The percentage of the modified area decreases quickly as the number of iterations increases.}
\label{fig_thresholdEstimation}
\end{figure}

According to observations, the algorithm iterates effectively, and the convergence of $\bf{L}$ is very fast even with an increased number of iterations. For example, in our experiments, dealing with a 512-by-512 image only needs about 10 iterations to reach a steady state. The generation of a specific QR code is shown in \figurename\ \ref{fig_thresholdEstimation}. As the number of iterations grows, the percentage of pixels that need updating decreases quickly, and the appearance of the QR code is accordingly stabilized.\par
\subsection{Color Aesthetic QR Codes}
After the generation of the grayscale QR code has been completed, it needs to be transformed into a color aesthetic one under the constraint of not changing the luminance. As can be inferred from Equation \ref{eqn_rgb2gray}, the mapping between RGB and grayscale is many-to-one, so there are many ways to convert three-color-channel images to single-channel grayscale ones.\par
During color aesthetic QR code generation, not only must the luminance be maintained, but the generated color image $\bf{Q^c}$ needs to resemble the original image $\bf{I}$ to guarantee the visual effect. The general method is to first convert $\bf{I}$ from RGB space to another color space (such as HSL, LAB), then to keep other channels from changing while only adjust the luminance channel until the corresponding grayscale image equals $\bf{Q^g}$, and finally convert the image back to RGB space. However, when converting the color space, the range of color expression is always limited, so simply adjusting luminance may not meet the expected requirements in some cases. In addition, luminance adjustment is usually transformed into an optimization problem, which increases the cost of calculation\cite{garateguy2014qr}.\par
\begin{figure}[!t]
\centering
\subfloat[Case I]{
\includegraphics[width=1.1in]{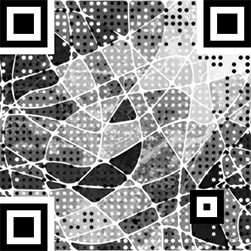}%
\quad
\includegraphics[width=1.1in]{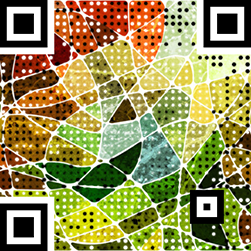}%
}
\hfil
\subfloat[Case II]{
\includegraphics[width=1.1in]{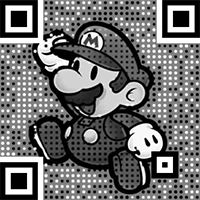}%
\quad
\includegraphics[width=1.1in]{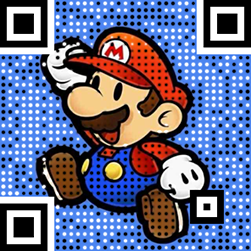}%
}
\caption{Example of grayscale-to-colored QR code conversion}
\label{fig_QRGray2RGB}
\end{figure}

Inspired by the bilinear interpolation algorithm, in this paper a linear-based luminance adjustment algorithm is proposed, which alleviates the above problems. In this method the process of luminance adjustment as is treated as a linear interpolation process. For convenience, we first define $C_m(x)$, which represents the minimum or maximum luminance value of position $x$. That is, when $ x\in M_k$,
\begin{equation}
\setlength{\nulldelimiterspace}{0pt}
C_m(x)=\left\{\begin{IEEEeqnarraybox}[\relax][c]{l?s}
{\left[255,255,255\right]}^{T},&for $Q^b_k=1$\\
{\left[0,0,0\right]}^{T},&for $Q^b_k=0$%
\end{IEEEeqnarraybox}\right.
\end{equation}
After that, according to the principles above, let
\begin{equation}
\label{eqn_lumCalcColor}
Q^c_x = I_x + \theta(C_m(x)-I_x)
\end{equation}
where $\theta$ is a variable, which is mainly used for luminance adjustment, whose range is from 0 to 1. While $Q^c_x$ is the output color value of the color image at $x$. Here the symbol $\omega$ will be used to denote ${\left[\alpha,\beta,\gamma \right]}^{T}$. As can be inferred from Equation \ref{eqn_rgb2gray}, in the process of scanning $\bf{Q^c}$, if the corresponding grayscale image is to match $\bf{Q^g}$, then
\begin{equation}
{Q^g}(x) = {\omega^T} Q^c_x
\end{equation}
Therefore, by multiplying both sides of Eq.~\ref{eqn_lumCalcColor} by $\omega^T$, we have:
\begin{equation}
{\omega^T}{Q^c_x} = {\omega^T}{I_x} + \theta{\omega^T}{(C_m(x)-I_x)}
\end{equation}
which, when combined with the previous equation, gives us:
\begin{equation}
\label{eqn_colorParameterCalc}
\theta = \frac{Q^g_x-{\omega^T}{I_x}}{{\omega^T}C_m(x)-{\omega^T}I_x}
\end{equation}
Combining Equation \ref{eqn_colorParameterCalc} and Equation \ref{eqn_lumCalcColor}, we finally have
\begin{equation}
Q^c_x = I_x + \frac{Q^g_x-{\omega^T}{I_x}}{{{\omega^T}C_m(x)-{\omega^T}I_x}} (C_m(x)-I_x)
\end{equation}
As can be seen from the above, all the variables on the right side are known, therefore it is easy to combine the grayscale aesthetic QR code $\bf{Q^g}$ with the original image $\bf{I}$, to generate a color aesthetic QR code $\bf{Q^c}$. Two examples of grayscale-to-colored QR codes are shown in \figurename\ \ref{fig_QRGray2RGB}.\par
\subsection{Parameter Analysis} \label{sec:parameterAnalysis}
Through the above analysis, a complete aesthetic QR code generation algorithm is proposed focusing on scanning robustness. The matrix ${\bm{\eta }}$ is used to ensure the scanning robustness of the generated code and ${\varpi}^p_x$ to initialize the weight of each module. However, the question remains regarding the choice of these values for actual applications, and their corresponding influence on the generation of aesthetic QR codes.\par
\begin{figure}[!t]
\includegraphics[width=3.5 in]{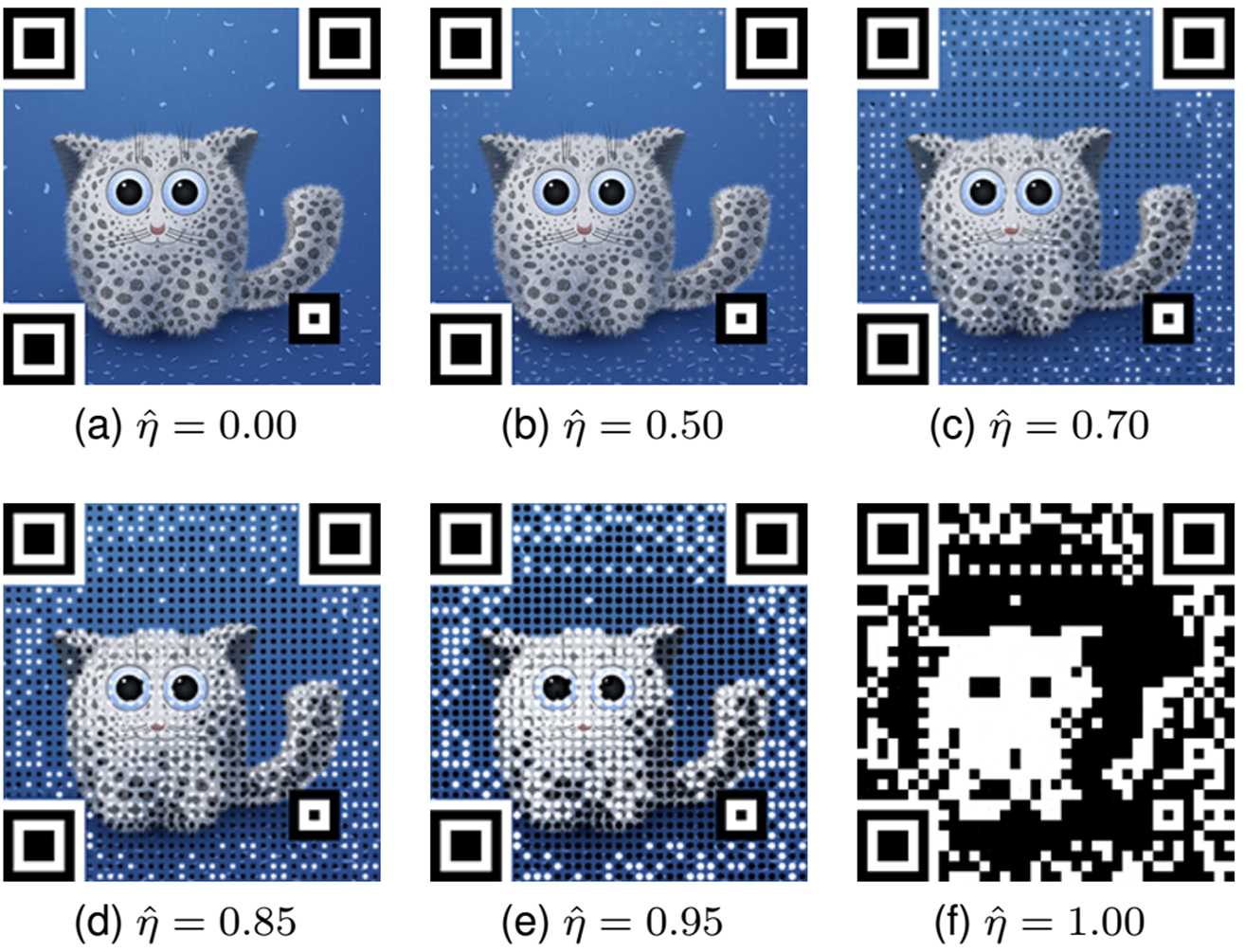}
\label{fig_sub_etaCurve}
\caption{Influence of ${\bm{\eta }}$ on the aesthetic QR code generation}
\label{fig_etaVariation}
\end{figure}

Firstly, to study the effect of ${\bm{\eta }}$ on the generated QR code, we conduct a set of experiments. We first investigate each element of ${\bm{\eta }}$ be equal, that is, ${{\eta}_k} = {\hat \eta }$. Then, a dataset containing 300 images was created, and aesthetic QR codes were generated for each value of ${\hat \eta }$ in the range 1.00, 0.99, ..., 0.00, respectively to create a total of 30300 QR code images. By scanning the resulting images, a curve which describes the relationship between the rate of scanning success and ${\hat \eta }$ was obtained. As shown in \figurename\ \ref{fig_sub_etaCurve}, with the decrease of ${\hat \eta }$, the rate of scanning success also declines, and gradually reaches 0.\par
In the generation algorithm of the aesthetic QR code, scanning robustness is mainly affected by ensuring that ${P_{M_k}}$, i.e. the probability of correctly scanning a single module, is not less than ${{\eta}_k}$. When it comes to thresholding, if each module is considered as an independent experiment of success or failure, the probability of successfully scanning the whole QR code is equal the percentage of instances where the failure rate is less than the necessary threshold for recovering the codeword. Therefore, improving the ${{\eta}_k}$ of a single module results in an increase of the overall scanning robustness of the QR codes.\par
As we observed in \figurename\ \ref{fig_etaVariation}a-f, the larger ${\hat \eta }$ is, the more similar the result image and the binary aesthetic QR code will be. When ${\hat \eta } = 1.00$, that is, ${{\eta}_k} = 1.00$ for all $k$, the generated color aesthetic QR code is exactly the same as the binary aesthetic QR code. On the contrary, the smaller ${\hat \eta }$ is, the more similar it will be to the original image. When ${\hat \eta }=0.00$, except for some functional pattern area of the generated color aesthetic QR code, the remaining areas resemble the original image exactly.\par
As can be inferred from the above discussion, for a single module, the larger ${{\eta}_k}$ is, the larger the probability of it being scanned correctly will be. In real applications, setting ${{\eta}_k} = {\hat \eta }$ for all $k$-s is not required, so ${{\eta}_k}$ is usually adjusted locally in order to obtain a better aesthetic effect. For instance, one can use
\begin{equation}
\bm{\eta} = 0.75 + 0.15(1-\bm{W})
\end{equation}
to keep ${\eta _k}$ in the range of $[0.75,0.9]$.\par
\begin{figure}[!t]
\centering
\includegraphics[width=3.4 in]{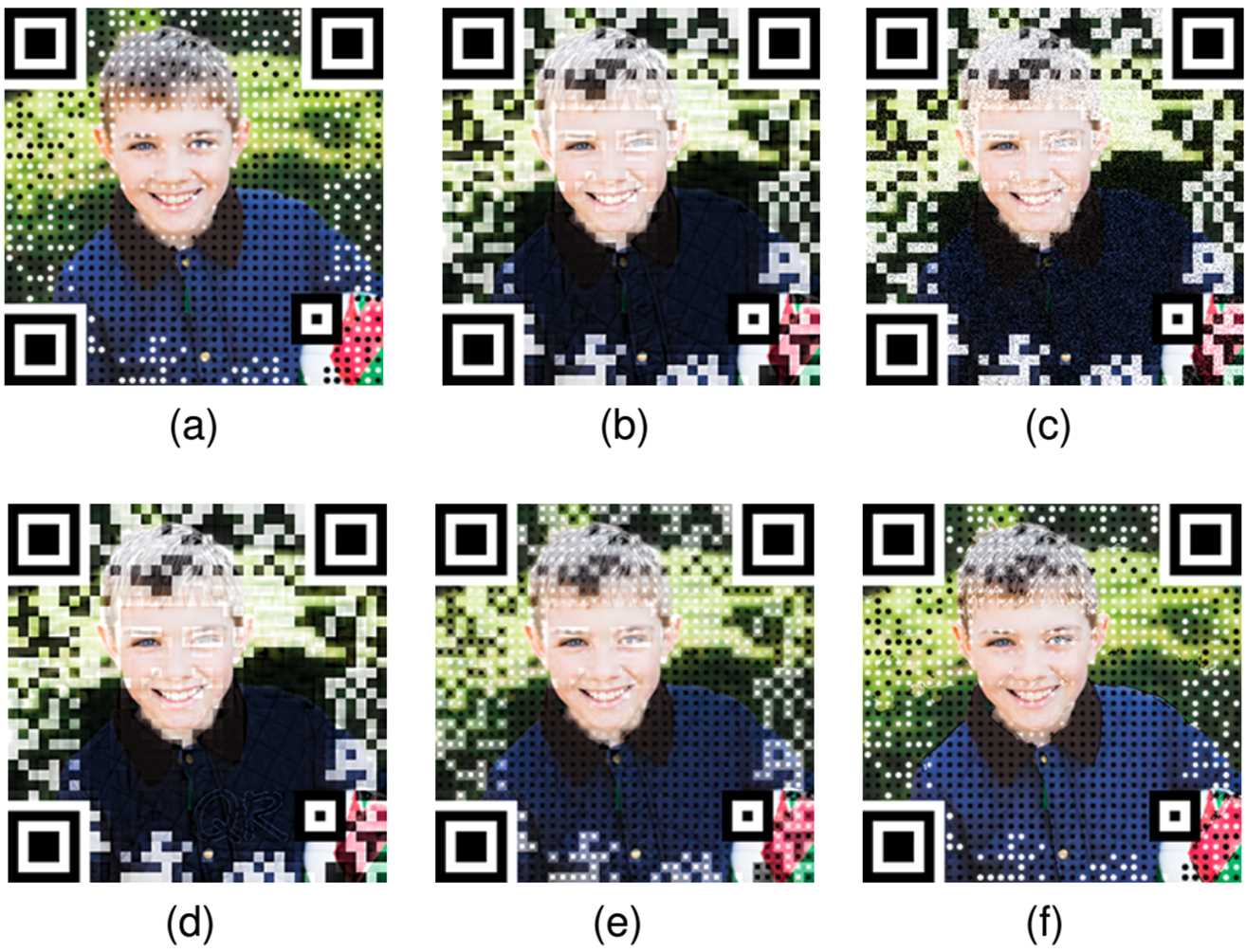}
\caption{Aesthetic QR codes of different styles using different initial ${\varpi}^p_x$ values. a. ${\varpi}^p_x$ using a 2-D Gaussian distribution; b. Constant value for all ${\varpi}^p_x$; c. initialization with a random in the range of [0,1]; d. Initialization using pixel values from another image, for example, to generate a patch of QR in the bottom-right of the image; e. Greater central than peripheral weights to formulating an image similar to Visualead's approach; f. Initialization using superposition of Gaussian matrix and image edge detection result.}
\label{fig_varpiVariation}
\end{figure}

Finally, the influence of ${\varpi}^p_x$ on the generated color aesthetic QR code is discussed. In Algorithm \ref{alg_adjustPointProbability}, it is used to initialize the range-adjusting weight of the module's probability. In the example mentioned above, it is treated as equal to $p^s_x$ by default. That is, as a 2-dimensional Gaussian distribution, in order to have a prior to the color adjustment of the module's center. However, in reality, more styles of color aesthetic QR code results can be obtained by using different values. \figurename\ \ref{fig_varpiVariation} shows some results of the same image for 6 different initializations of ${\varpi}^p_x$.
The scanning robustness can still be maintained for different appearances, which indicates that the constraint of ${\bm{\eta }}$ is effective.\par
The use of the variables (${\bm{\eta }}$ and ${\varpi}^p_x$) provides flexibility to the generation of color aesthetic QR, which enables users to generate aesthetic QR codes of different robustness and visual styles according to their requirements.\par
\section{Experiments}
In this section, the visual effects and the robustness of QR codes are evaluated through extensive experiments. First of all, a user survey was conducted to compare the visual effects of different methods. After that, the robustness of QR codes was quantified and compared through decoding experiments and user evaluations.\par
\subsection{Visual Effects}
To evaluate the visual effect of color aesthetic QR code, a user survey was conducted. Different types of images were first chosen, including cartoons, sceneries, buildings, animals, human faces, brands, etc. After that, 4 different methods of generating aesthetic QR codes were applied to generate the corresponding code for each image. Among them, Halftone\cite{chu2013halftone} is a popular generation method of aesthetic QR codes. Visualead\cite{visualead} is a widely used commercial algorithm, and Efficient\cite{linefficient} is the latest QR code visual beautification algorithm which has only recently been presented. We use these methods to obtain 20 groups of images in total, each group containing an original image and 4 QR code images generated using different methods. In this way, an evaluation questionnaire for the visual effect was created. The sample images are shown in Table \ref{table_aesthetic_QRCode}.\par
\begin{table*}[!t]
\caption{Examples Generated using Different Algorithms}
\centering
\begin{tabular}{p{7in}}
\mgape{\includegraphics{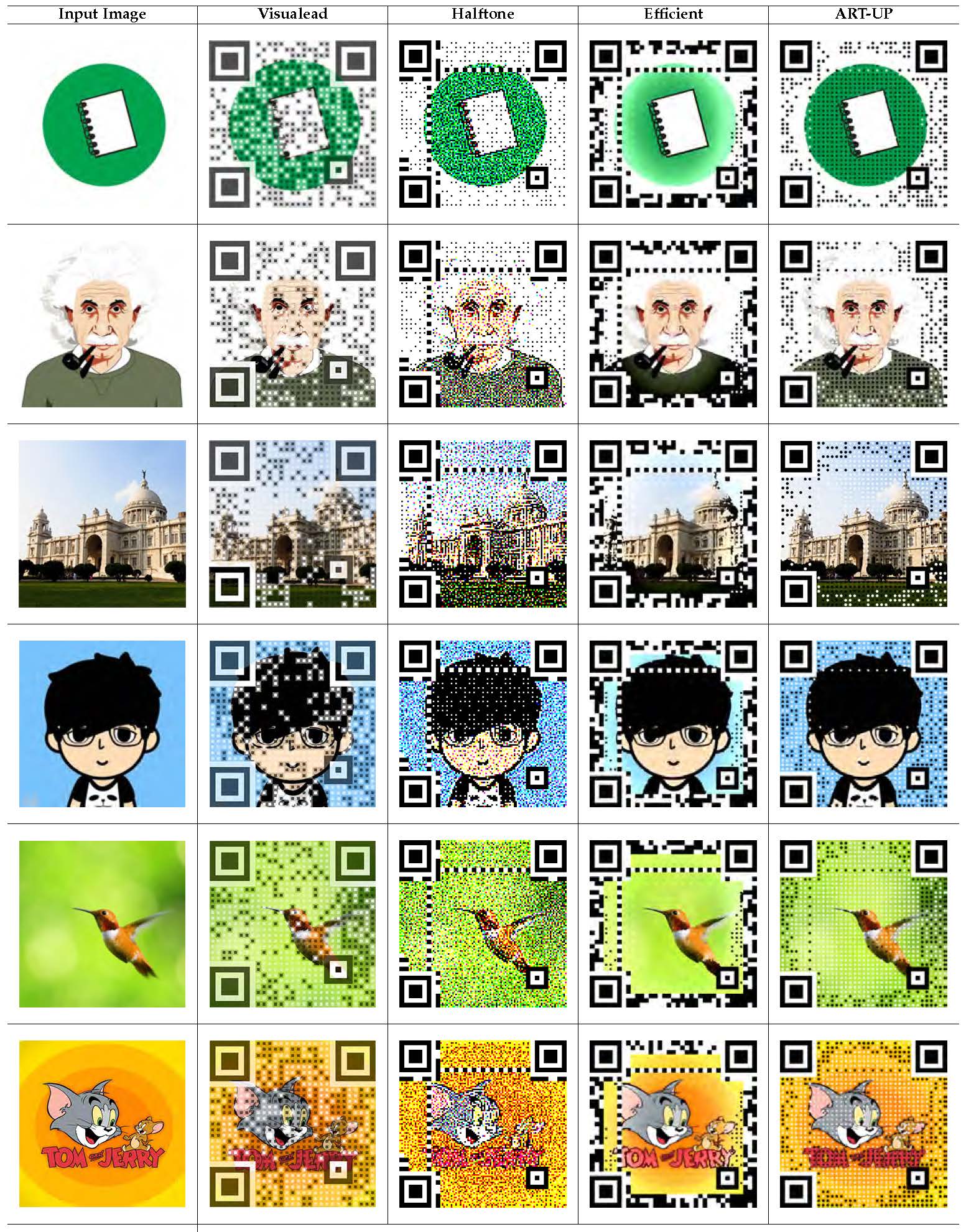}}
\end{tabular}
\label{table_aesthetic_QRCode}
\end{table*}
\begin{figure}[!t]
\centering
\subfloat[Aesthetic Measure]{
\includegraphics[width=1.5in]{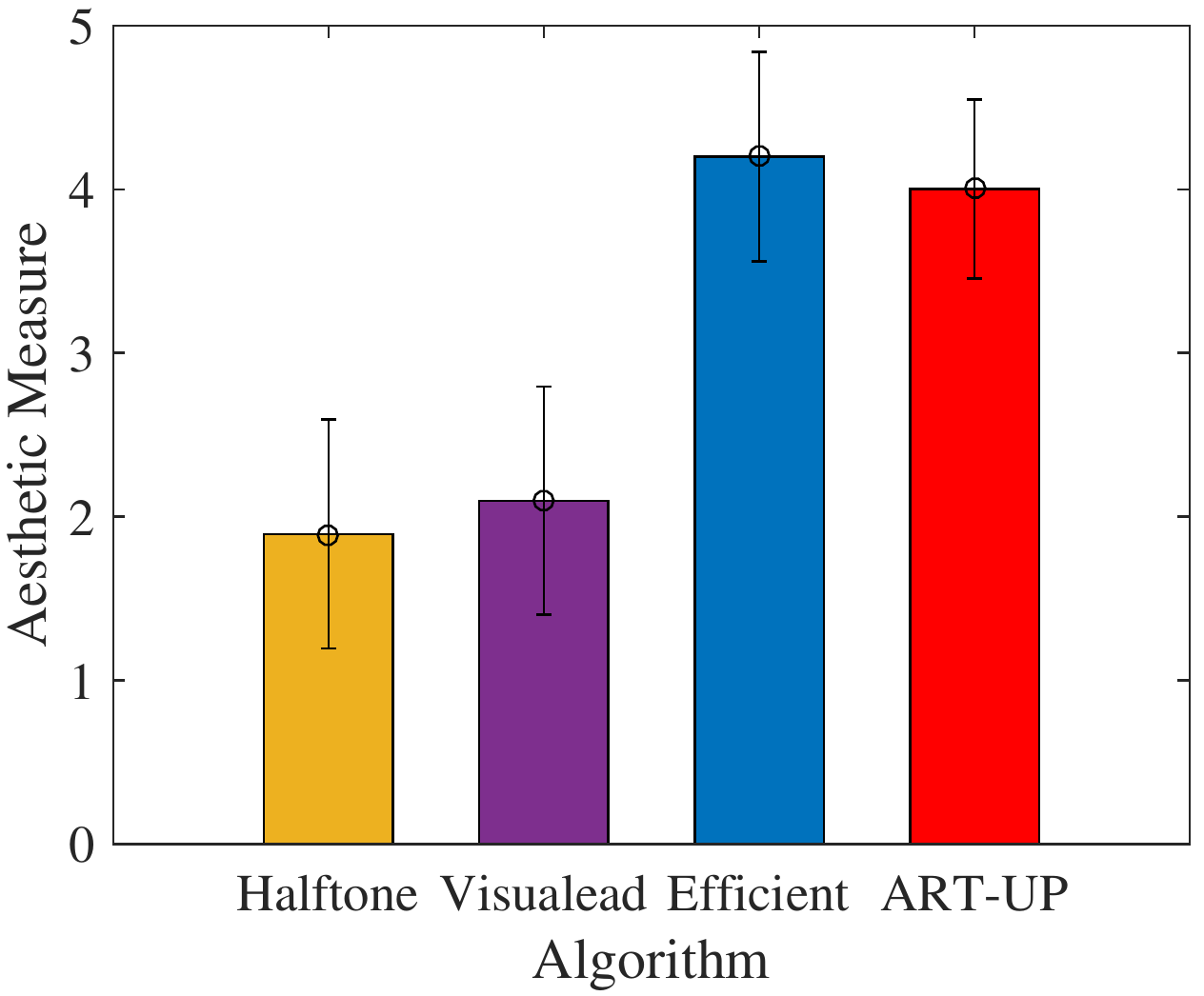}%
}
\subfloat[Similarity Measure]{
\includegraphics[width=1.5in]{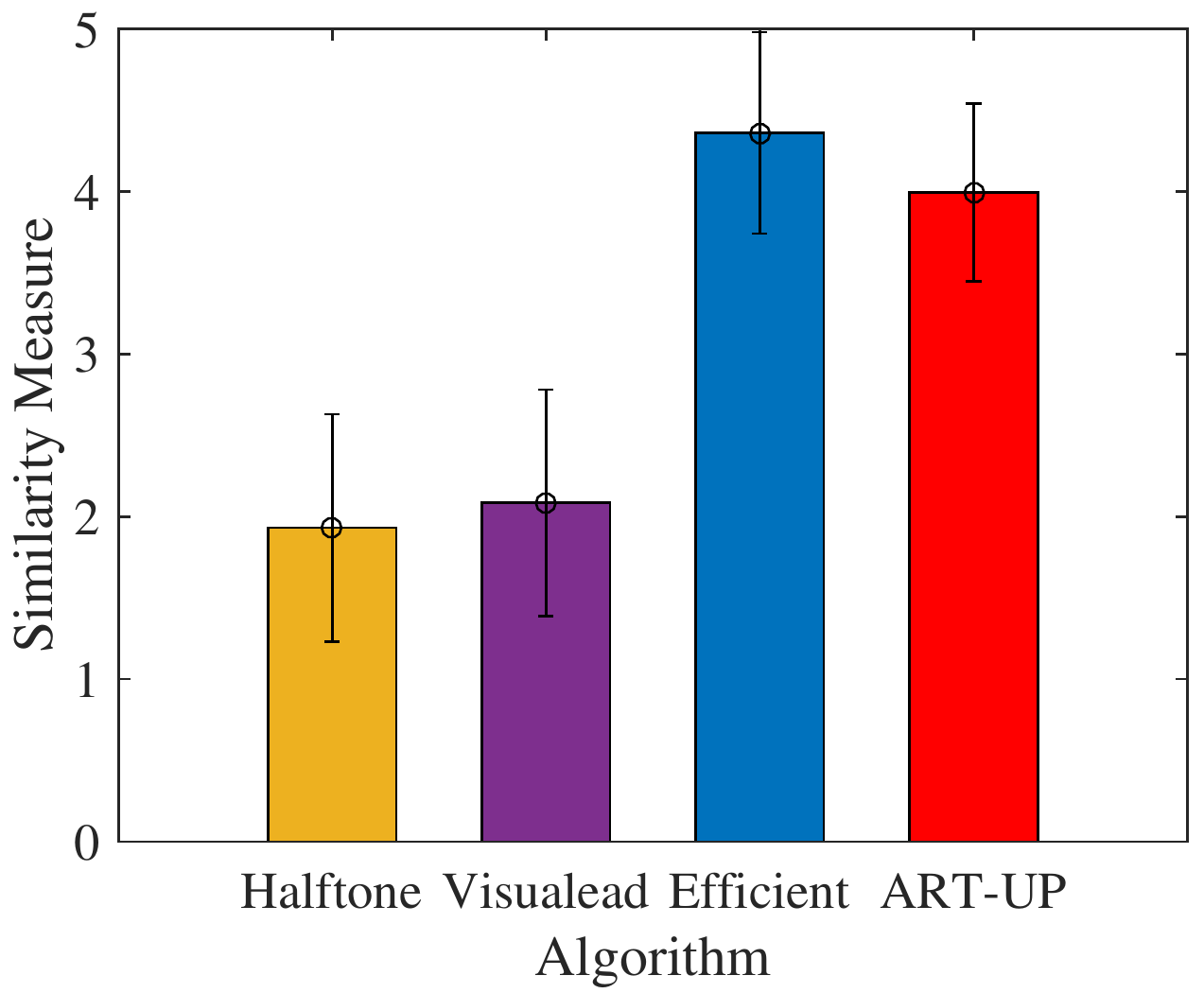}%
}
\caption{User study about aesthetic and similarity}
\label{fig_userStudyResult}
\end{figure}
40 volunteers, 25 males and 15 females, were invited to participate in the survey. The images of each group were displayed, they were asked to give subjective scores ranging from 1 to 5 according to the observed degree of visual appeal and similarity to the original images.\par
As shown in \figurename\ \ref{fig_userStudyResult}, the visual effect of the QR code images generated by the four algorithms and the average scores of the similarities between the generated images and the original ones were evaluated. \par
It can be seen that the scores of the two measures are similar, and the visual effect generated by ART-UP which this paper proposes resembles the state-of-art Efficient method \cite{linefficient}. Besides, it behaves much better than the Visualead \cite{visualead} and Halftone \cite{chu2013halftone} algorithms.\par
\begin{figure}[!t]
\centering
\includegraphics[width=1.8in]{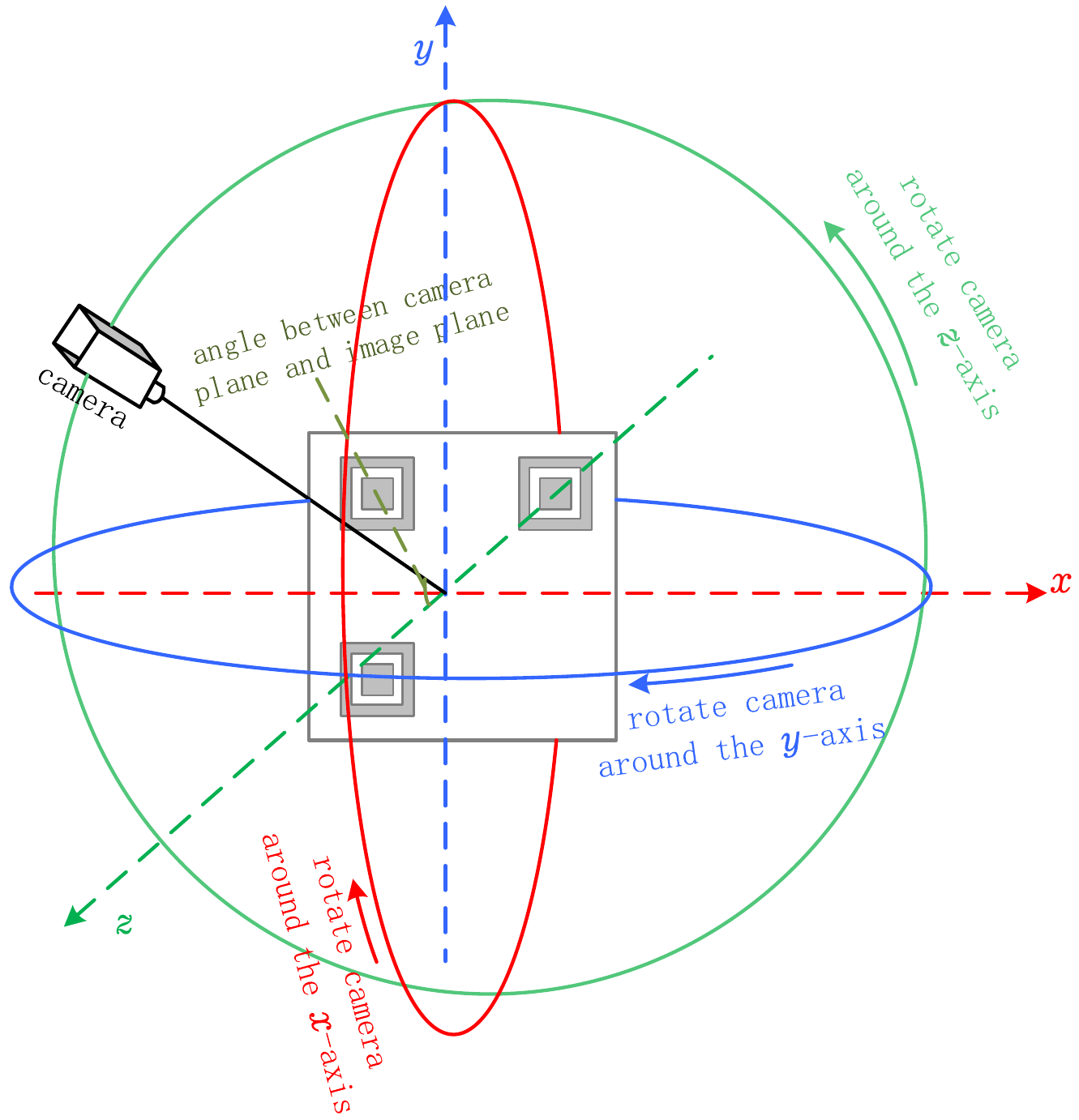}
\caption{The test method for scanning robustness of angle variation. The QR code was placed in the center of the coordinate system, with the camera a fixed distance from the origin and aimed toward the center.}
\label{fig_rotateXYZ}
\end{figure}
\begin{figure*}[!t]
\centering
\includegraphics[width=7in]{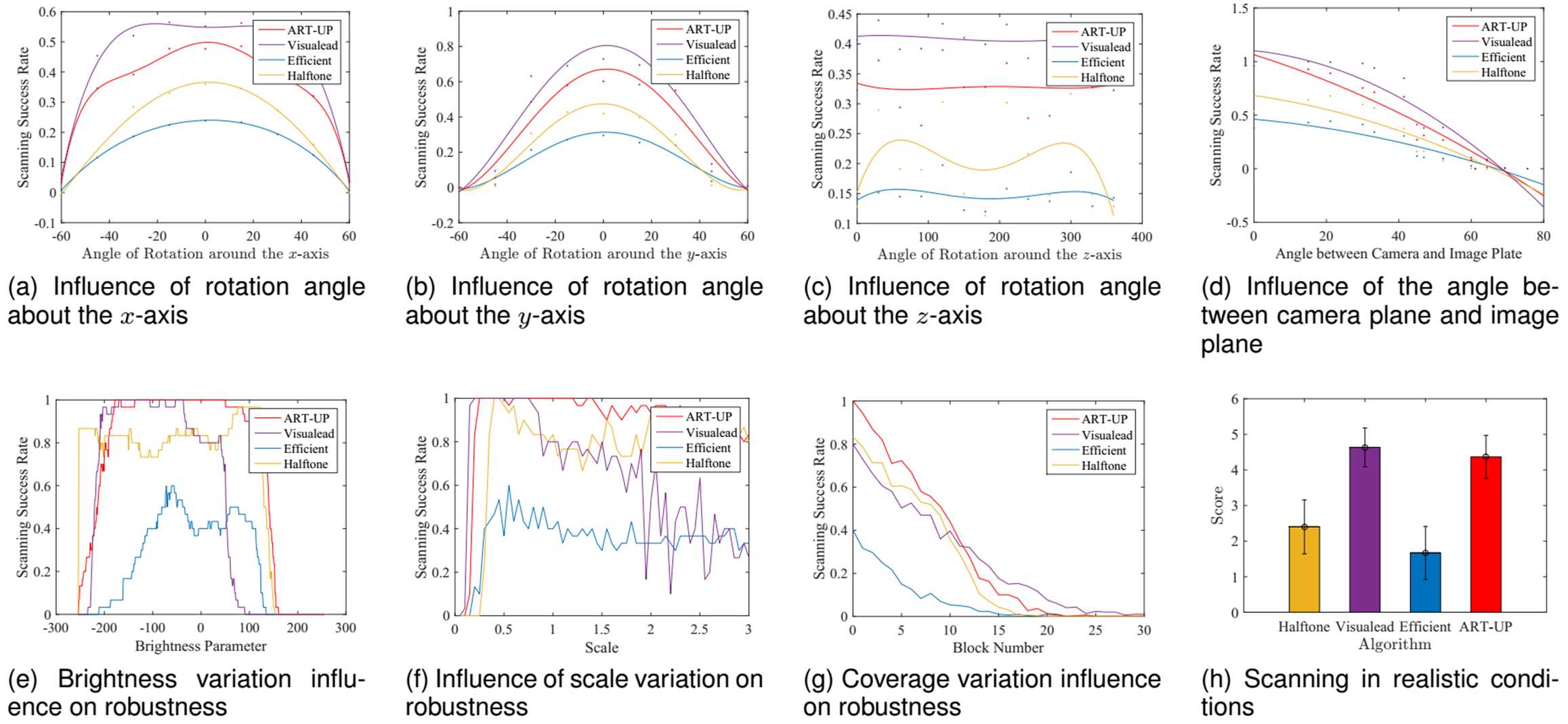}
\label{fig_evaluation}
\end{figure*}
\subsection{Scanning Robustness}
Scanning robustness of the QR codes was evaluated from various aspects, such as scanning angle variation, brightness variation, scale variation, coverage, and scanning in real environments. 30 images were randomly selected from the dataset to establish a sub-dataset from the 300-image dataset, and the robustness tests were carried out on this subset.\par
\emph{Scanning Angle Variation.} QR code images were placed in the origin of a 3-D axes system, and the camera was rotated about the $x$-, $y$- and $z$-axes respectively, as shown in \figurename\ \ref{fig_rotateXYZ}. For the $x$- and $y$-axes, the images were sampled in $15^\circ$ steps ranging from $-60^\circ$ to $60^\circ$. For the $z$-axes, the images were sampled in $30^\circ$ steps ranging from $0^\circ$ to $360^\circ$. In this way, 972 measurements were obtained. The corresponding results are shown in \figurename\ \ref{fig_sub_x_variation},\ref{fig_sub_y_variation},\ref{fig_sub_z_variation}, respectively. Meanwhile, the angle between the plane of the QR code and the one of the camera was also calculated and the corresponding results are shown in \figurename\ \ref{fig_sub_plate_variation}. As can be inferred from the four figures, ART-UP is slightly lower than that of Visualead's state-of-the-art commercial algorithm in terms of scanning angle robustness, but better than the other two methods.\par
\emph{Brightness Variation.} QR code scanning is easily affected by illumination and capturing devices, resulting in differences between the captured and the real luminance. To study this effect further, it was simulated with a linear brightness adjustment, adjusting the brightness of each image varying $-255$ to $+255$. That is, each QR code image results in 511 images of different luminance levels. As shown in \figurename\ \ref{fig_sub_brightness_variation}, with regard to small-range brightness change, ART-UP is obviously more robust and immune to illumination changes among all methods.\par
\emph{Scale Variation.} For aesthetic QR codes, scale variation is always a key factor that affects scanning. Whether printed or shown on a screen, QR codes face scaling issues. For each image in the dataset, a $512 \times 512$ QR code image was generated, and was sampled at a $0.05$ step per frame for a scaling ratio ranging from $0.05$ to $3.00$. That is, each QR code image yielded 61 images at different scaling ratios. The image scaling was performed through bi-cubic interpolation. As shown in \figurename\ \ref{fig_sub_size_variation}, ART-UP demonstrates the best performance compared to the other three schemes when faced with the scale variation and it is consistent. Efficient is consistent but has a low scanning success rate, while Visualead shows serious oscillation and inconsistency.\par
\emph{Coverage.} QR codes allow a considerable percentage of error correction, which will be affected when the code is covered or has a missing segment. Each QR code of the image was replaced with a fixed number of $2a \times 2a$ square black-or-white blocks in random positions. After 30 rounds of repeated testing, the average performance was determined. As shown in \figurename\ \ref{fig_sub_cover_variation}, ART-UP, Visualead, and Halftone show similar performance, while the Efficient algorithm suffers more in these cases.\par
\emph{Scanning in realistic conditions.} Besides simulating scanning experiments, one subjective experiment was carried out, where 30 volunteers were invited to vote for the QR codes' performance under real scanning circumstances. In the experiment, no constraint on the scanning tools, scanning environments or scanning methods was set. Volunteers were shown the QR codes generated by the different algorithms, and were required to scan each one. The subjective score ranged from 1 to 5 regarding the response time, where a score of 1 indicates a scarce ability to successfully scan the code and the score of 5 represents a prompt response. As shown in \figurename\ \ref{fig_sub_real_variation}, the result of ART-UP is almost equal to the Visualead's, and both of them are ranked better than Halftone and Efficient. This experiment demonstrates that the proposed algorithm is robust enough to adapt to most common scenes. Especially when the codes are printed on paper or other materials, the speed of response is still quick and stable.\par

\section{Conclusion}
In this paper, a flexible generation algorithm of aesthetic QR codes is proposed, where a QR code is obtained based on a probability model of successfully scanning the QR codes, controlling and adjusting the tradeoff between visual effect and scanning robustness. By adjusting the thresholding and sampling errors, our proposed algorithm can generate aesthetic QR codes with different appearances using a hierarchical, coarse-to-fine strategy. Experimental results show that ART-UP is scanning-robust and can avoid the error occurrence in most realistic scenarios. This algorithm not only provides aesthetic results, but also ensures code usability, and is thus suitable for commercial needs.\par





\ifCLASSOPTIONcompsoc
\else
\fi



\bibliographystyle{IEEEtran}
%
\bibliography{IEEEabrv,QRCode}

%

%

\begin{IEEEbiography}[{\includegraphics[width=1in,height=1.25in,clip,keepaspectratio]{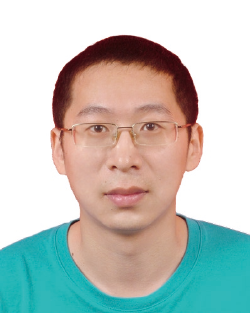}}]{Mingliang Xu}
is an associate professor in the School of Information Engineering of Zhengzhou University, China, and currently is the director of CIISR ( Center for Interdisciplinary Information Science Research), and the general secretary of ACM SIGAI China. His research interests include virtual reality and artificial intelligence. Xu got his Ph.D. degree in computer science and technology from the State Key Lab of CAD\&CG at Zhejiang University.
\end{IEEEbiography}
\vspace{-1.1 cm}
\begin{IEEEbiography}[{\includegraphics[width=1in,height=1.25in,clip,keepaspectratio]{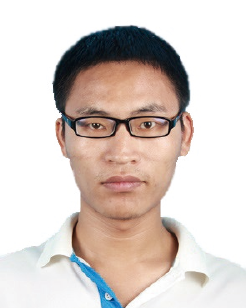}}]{Qingfeng Li}
is a master student in Center for Interdisciplinary Information Science Research, Zhengzhou University, China. He received the B.E. degree in Computer Software from Zhengzhou University, in 2014. He is currently working with Prof. Jianwei Niu in Beihang University, his research interests include computer version, image processing, and computer graphics.
\end{IEEEbiography}
\vspace{-1.1 cm}
\begin{IEEEbiography}[{\includegraphics[width=1in,height=1.25in,clip,keepaspectratio]{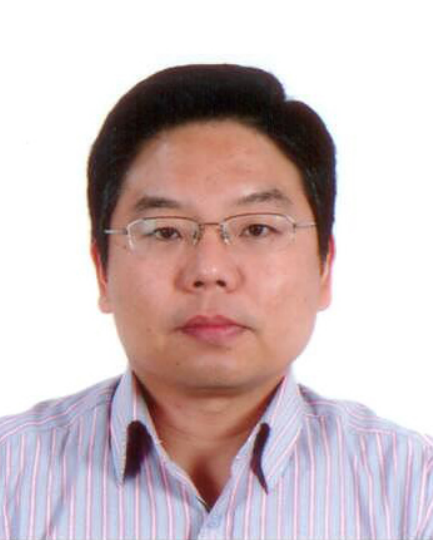}}]{Jianwei Niu}
received the M.S. and Ph.D. degrees in computer science from Beihang University, Beijing, China, in 1998 and 2002, respectively. He was a visiting scholar at School of Computer Science, Carnegie Mellon University, USA from Jan. 2010 to Feb. 2011. He is a professor in the School of Computer Science and Engineering, BUAA, and an IEEE senior member. His current research interests include mobile and pervasive computing, mobile video analysis.
\end{IEEEbiography}
\vspace{-0.9 cm}

\begin{IEEEbiography}[{\includegraphics[width=1in,height=1.25in,clip,keepaspectratio]{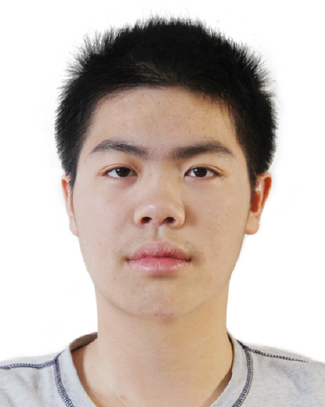}}]{Xiting Liu}
is currently a senior student in Beihang University, Beijing, China. His major is Computer science and technology. He once studied and worked with Prof. Jianwei Niu, in 2016. His research interests include computer version and image processing.
\end{IEEEbiography}
\vspace{-1.1 cm}

\begin{IEEEbiography}[{\includegraphics[width=1in,height=1.25in,clip,keepaspectratio]{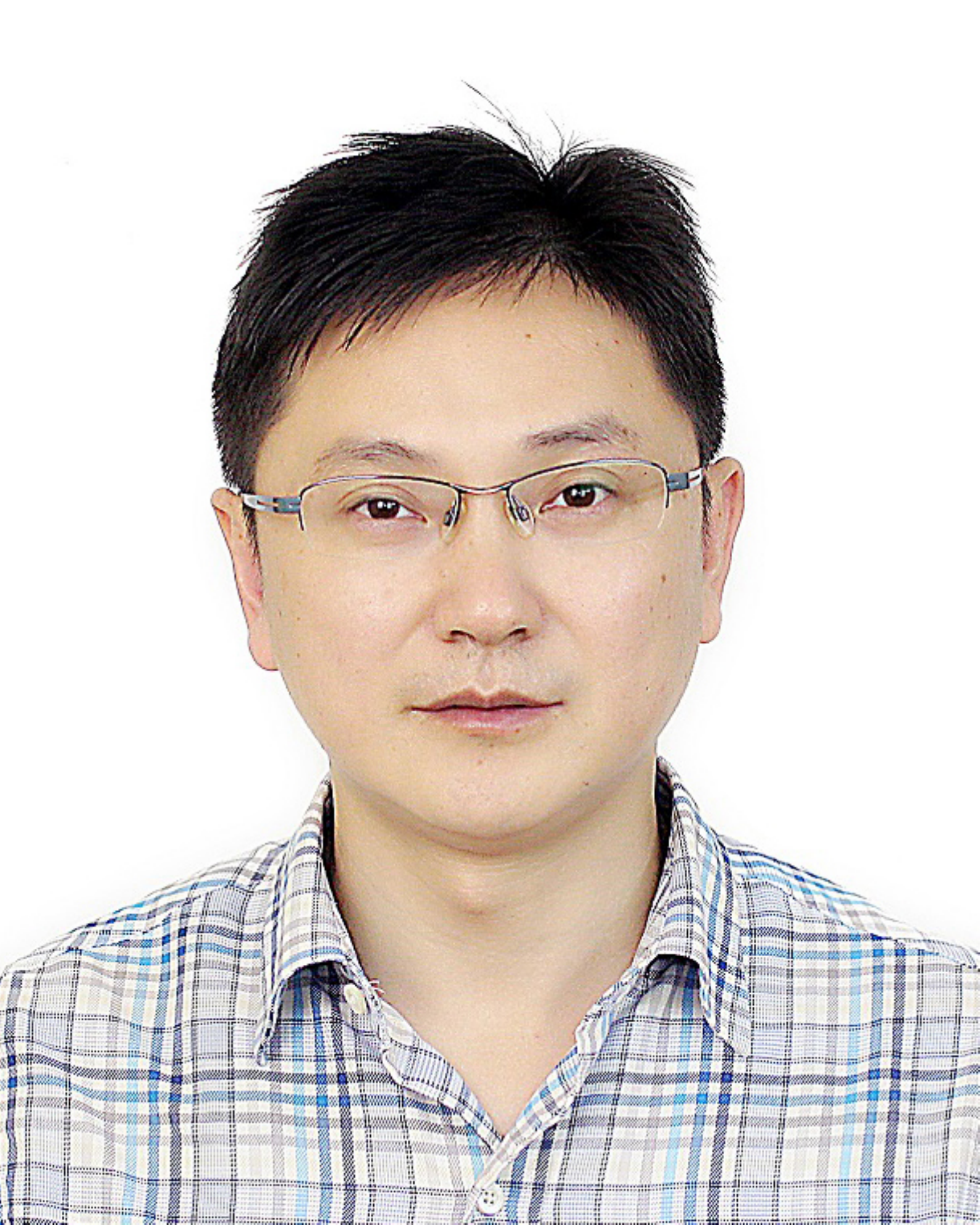}}]{Weiwei Xu}
now is  a researcher in State Key Lab of CAD\&CG, College of Computer Science at Zhejiang University, recipient of the NSFC Excellent Young Scholars
Program in 2013. His main research interests are digital geometry processing, physical simulation and virtual reality.
\end{IEEEbiography}
\vspace{-1.1 cm}
\begin{IEEEbiography}[{\includegraphics[width=1in,height=1.25in,clip,keepaspectratio]{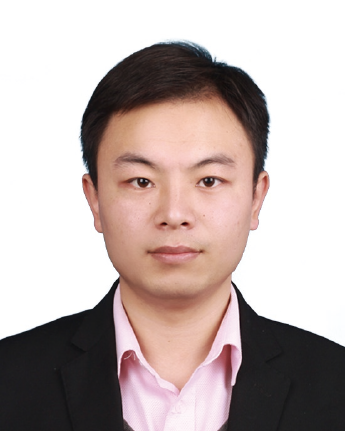}}]{Pei Lv}
is an assistant professor in Center for Interdisciplinary Information Science Research, Zhengzhou University, China.His research interests include video analysis and crowd simulation. He received his Ph.D in 2013 from the State Key Lab of CAD\&CG, Zhejiang University, China.
\end{IEEEbiography}

\vspace{-1.1 cm}
\begin{IEEEbiography}[{\includegraphics[width=1in,height=1.25in,clip,keepaspectratio]{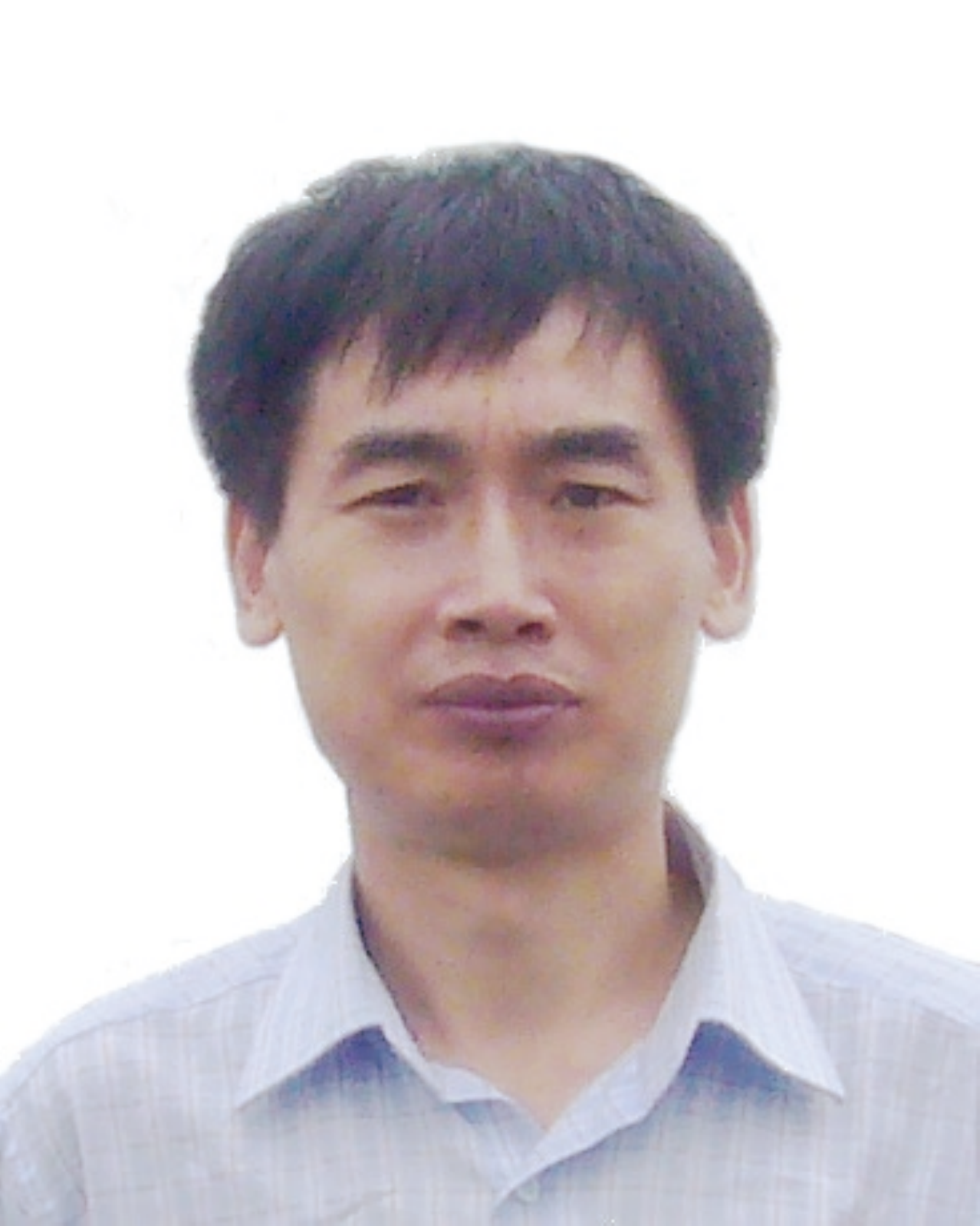}}]{Bing Zhou}
is currently a professor in Center for Interdisciplinary Information Science Research, Zhengzhou University, Henan, China. He received the B.S. and M.S. degrees from Xi’an Jiaotong University in 1986 and 1989, respectively,and the Ph.D. degree in Beihang University in 2003, all in computer science. His research interests cover video processing and understanding, surveillance, computer vision, multimedia applications.
\end{IEEEbiography}





\end{document}